\documentclass[lettersize,journal]{IEEEtran}
\usepackage{amsmath,amsfonts}
\usepackage{xcolor} 
\usepackage{hyperref}
\usepackage{array}
\usepackage{booktabs} % For formal tables
\usepackage{multirow}
\usepackage{siunitx} % For table spacing
\usepackage{textcomp}
\usepackage{stfloats}
\usepackage{url}
\usepackage{booktabs} % For better looking tables
\usepackage{siunitx}  % Allows for alignment and formatting of numbers in tables
\usepackage{multirow} % Allows for multirow cells in tables
\usepackage{color}    % To use color
\usepackage{verbatim}
\usepackage{graphicx}
\usepackage{cite}
\usepackage{graphicx}
\usepackage{enumitem,kantlipsum}
\usepackage{subcaption}
\usepackage{tablefootnote}
\usepackage{cleveref}
\usepackage{svg}
\usepackage{textcomp}
\usepackage{breqn}
\usepackage{tikz}
\usepackage{adjustbox}
\usepackage{multirow}
\usepackage{xcolor}
\usepackage{pgfplots}
\usepackage{booktabs}
\usepackage[vlined, ruled, lined, longend, linesnumbered]{algorithm2e}
\usepackage[utf8]{inputenc}
\usepackage{tikz}
\usepackage{amssymb}% http://ctan.org/pkg/amssymb
\usepackage{pifont}% http://ctan.org/pkg/pifont
\newcommand{\xmark}{\ding{55}}%
\begin{document}

%\title{Zero-X: Securing IoV from Zero-Day Attacks with Blockchain and Open-set Federated Learning}

\title{Zero-X: A Blockchain-Enabled Open-Set Federated Learning Framework for Zero-Day Attack Detection in IoV}

\author{Abdelaziz Amara korba, \textit{Member}, \textit{IEEE}, Abdelwahab Boualouache, \textit{Member}, \textit{IEEE}, and Yacine Ghamri-Doudane, \textit{Senior Member}, \textit{IEEE}

\thanks{A. Amara korba is with the L3I, University of La Rochelle, France, and also with the LRS, Badji Mokhtar Annaba University, Algeria (e-mail : abdelaziz.amara\_korba@univ-lr.fr). Abdelwahab Boualouache is with FSTM, University of Luxembourg, Luxembourg (e-mail: abdelwahab.boualouache@uni.lu). Y. Ghamri-Doudane is with the L3I, University of La Rochelle, France (e-mail : yacine.ghamri@univ-lr.fr)}
}

% \author{\IEEEauthorblockN{1\textsuperscript{st} Abdelaziz {Amara korba}}
% \IEEEauthorblockA{\textit{L3I, University of La Rochelle, France } \\
% \textit{LRS, Badji Mokhtar Annaba University, Algeria}\\
% %La Rochelle, France \\
% %email address or ORCID
% }
% \and
% \IEEEauthorblockN{2\textsuperscript{nd} Abdelwahab Boualouache}
% \IEEEauthorblockA{\textit{FSTM, University of Luxembourg} \\
% %\textit{name of organization (of Aff.)}\\
% Luxembourg\\
% %email address or ORCID
% }

% \and
% \IEEEauthorblockN{3\textsuperscript{rd} Yacine Ghamri-Doudane}
% \IEEEauthorblockA{\textit{L3I, University of La Rochelle,France} \\
% %\textit{name of organization (of Aff.)}\\
% % \\
% %email address or ORCID
% }
% }


\markboth{}%
{Shell \MakeLowercase{\textit{et al.}}: A Sample Article Using IEEEtran.cls for IEEE Journals}


\maketitle

\begin{abstract}
The Internet of Vehicles (IoV) is a crucial technology for Intelligent Transportation Systems (ITS) that integrates vehicles with the Internet and other entities. The emergence of 5G and the forthcoming 6G networks presents an enormous potential to transform the IoV by enabling ultra-reliable, low-latency, and high-bandwidth communications. Nevertheless, as connectivity expands, cybersecurity threats have become a significant concern. The issue has been further exacerbated by the rising number of zero-day (0-day) attacks, which can exploit unknown vulnerabilities and bypass existing Intrusion Detection Systems (IDSs). In this paper, we propose Zero-X, an innovative security framework that effectively detects both 0-day and N-day attacks. The framework achieves this by combining deep neural networks with Open-Set Recognition (OSR). Our approach introduces a novel scheme that uses blockchain technology to facilitate trusted and decentralized federated learning (FL) of the Zero-X framework. This scheme also prioritizes privacy preservation, enabling both CAVs and Security Operation Centers (SOCs) to contribute their unique knowledge while protecting the privacy of their sensitive data. To the best of our knowledge, this is the first work to leverage OSR in combination with privacy-preserving FL to identify both 0-day and N-day attacks in the realm of IoV. The in-depth experiments on two recent network traffic datasets show that the proposed framework achieved a high detection rate while minimizing the false positive rate. Comparison with related work showed that the Zero-X framework outperforms existing solutions.
\end{abstract}


\section{Introduction}
With the rapid development of computing and communication technologies, we are witnessing a growing prevalence of Connected and Automated Vehicles (CAVs) in our modern world. Internet of Vehicles (IoV) technology provides a crucial communication framework for CAVs, facilitating reliable communication between CAVs and other IoV entities, such as infrastructure, pedestrians, and smart devices. The fifth generation (5G) and the forthcoming sixth generation (6G) networks promise to revolutionize the IoV by enabling ultra-reliability with ultra-low latency and high bandwidth communications.

%Zero-day & IoV
With the growing level of connectivity and complexity in the IoV, cybersecurity risks have become a major concern. Among these risks, Zero-day cyberattacks stand out as a significant threat. A zero-day attack refers to a new type of cyber attack that is unknown to both the general public and the cybersecurity experts \cite{guo2022review}. This kind of attack exploits unknown vulnerabilities or uses innovative methods to evade detection by security mechanisms. Upstream \cite{turgeman_2022}, a cybersecurity and data management platform for CAVs has analyzed over 900 publicly reported cyberattacks on cars in the last decade. Examples of real-world cyberattacks can be found in \cite{turgeman_2022}, including instances where hackers could disable the brakes and kill a car's engine traveling at 65 mph using a laptop and custom-written software plugged into the OBD II port. In another case, hackers remotely controlled a Jeep's engine while driving on a highway. These incidents highlight the significant cybersecurity risks associated with IoV and the urgent need for robust cybersecurity measures to ensure the safety of vehicle occupants and other road users.

% Limit of closed-set and supervised learning 
To mitigate cyberattacks in the context of the IoV, Artificial Intelligence (AI) has emerged as a crucial tool for cybersecurity. Machine Learning (ML) and Deep Learning (DL) based IDSs have been proposed to protect vehicular networks from cyber-attacks \cite{boualouache2022survey}. In ML, classification involves assigning data to predefined categories based on their features, using a trained model. This model, developed with a labeled dataset, learns to recognize patterns indicative of each class for predicting unlabeled data. In a real IoV environment, zero-day attacks will inevitably occur frequently. However, existing IDSs \cite{abdel2021federated,rahal2022antibotv,boualouache2022federated } are typically designed for static and closed-set scenarios. Assuming that all possible attack types an intrusion can belong to are known and predefined during the training phase of the IDS.  However, this assumption is unrealistic as new types of attacks are constantly emerging. The majority of current IDSs rely on Supervised Learning (SL), a method that can prove to be inefficient in effectively identifying unknown attacks characterized by patterns substantially divergent from those observed during the training phase.

To enable the detection of zero-day attacks, anomaly and novelty detection techniques have been used \cite{korba2023federated,jeong2023x, ashraf2020novel}. Although the solutions based on these techniques yield promising results in detecting unseen/ zero-day attacks, they can unfortunately not recognize the type of detected attacks. This poses a substantial challenge in the development of a security mechanism capable of identifying known (N-day) attacks while also detecting emerging zero-day (0-day) attacks.

Open-Set Recognition (OSR) \cite{geng2020recent} offers a realistic approach in a dynamic field like intrusion detection. This method equips models to manage both known and unexpected attack types, considering the limitations of incomplete training data and the possibility of encountering new attack types after training. OSR emphasizes the need for classifiers to be both accurate and flexible, adapting to scenarios where test samples may include previously unseen attack types. This adaptability is crucial for effectively responding to the continuously evolving nature of such fields. In recent years, combining deep neural networks with open-set recognition has led to significant advances in detecting unknown classes in computer vision \cite{kong2021opengan, neal2018open, oza2019c2ae}. However, this approach has yet to be fully utilized in intrusion detection. While some studies have explored the application of open-set recognition in this context \cite{zhang2022unknown, 9643172}, to the best of our knowledge, OSR has never been leveraged to detect 0-day attacks in IoV.

%Limit of centralized training
Recent IDSs ~\cite{uprety2021privacy,boualouache2022federated,liu2021blockchain,hbaieb2022federated} leveraged the potential of Federated Leaning (FL) paradigm to train the detection model in a distributed and privacy-preserving manner. FL enables collaborative training of the model without requiring the transmission of raw data, which enhances participants' data privacy. Due to centralized model aggregation, however, standard FL is susceptible to server failures and external attacks, which can result in inaccurate detection model updates or training failures. 

% Our solution 
To address the limitations mentioned in existing IDSs within the IoV environment, this paper introduces the Zero-X framework. This security framework effectively identifies both 0-day and N-day attacks by integrating deep neural networks with Open-Set Recognition (OSR). Our approach introduces a pioneering scheme that utilizes blockchain technology to facilitate secure and decentralized federated training of the Zero-X framework. The framework not only guarantees high accuracy but also ensures fast detection, which is a critical factor in mitigating the impact of attacks. The Zero-X framework is designed to detect and classify attacks solely based on network traffic. Its device-agnostic nature makes it suitable for deployment on both CAVs and Multi-access Edge Computing (MEC) infrastructure. Its purpose is to secure CAVs against inter-vehicular attacks while protecting the MEC infrastructure from possible attacks, including Distributed Denial of Service (DDoS) attacks, that may originate from compromised CAVs. In summary, the main contributions of this paper are:
\begin{itemize}

\item Our proposed framework uses unsupervised privacy-preserving FL to train an attack detection (AD) model. Specifically, we use a Deep Auto-Encoder (DAE) to model the expected communication pattern of CAVs and detect any deviations from this pattern as malicious activity. This approach is highly effective in detecting various types of attacks, as opposed to existing methods that require separate training for each new attack.
  
\item The Zero-X framework leverages open-set federated learning to train the attack classifier (AC) model, a deep multi-class data descriptor that aims to identify a spherical decision boundary for each type of attack. This boundary determines whether a network flow belongs to a given attack type, making it effective in detecting unseen attacks while accurately identifying known attacks. To the best of our knowledge, this is the first work to leverage OSR in combination with FL for intrusion detection in the context of the IoV.

\item We propose a new training scheme utilizing blockchain technology to empower the federated learning, enhancing the security and decentralization of the Zero-X framework's training process. Our contribution lies in the introduction of an innovative Byzantine Fault Tolerance consensus mechanism named Proof-of-Accuracy (PoA). This mechanism plays a pivotal role in guaranteeing the secure dissemination and aggregation of FL model updates.

\item The framework's effectiveness is thoroughly assessed by extensive evaluations on two recent datasets. The first dataset, 5G-NIDD~\cite{5GNIDD}, comprises 5G network traffic traces of attacks that target the infrastructure (MEC). The second dataset, VDoS~\cite{VDoS}, contains network traffic originating from inter-vehiclar attacks. To create a realistic test scenario, we designate one type of
attack as the 0-day attack, while the remaining attack types are
considered as N-day attacks.This process is repeated with
different 0-day attacks, resulting in $K$ test scenarios per dataset, where $K$ represents the number of attack types in the dataset.
\end{itemize}

\color{black}

The remainder of this paper is outlined as follows. Section~\ref{RT} provides an overview of the related work. Section \ref{SD} introduces the system design of the framework. In Section~\ref{SOL}, we present in detail the development and operation of the framework. The performance evaluation results are depicted in Section~\ref{SIM}. Finally, Section~\ref{CON} concludes the paper.
% \usepackage{booktabs}

\section{Related Work} \label{RT}
The development of IDS for IoV has gained significant attention in recent years due to its criticality in ensuring the security and privacy of these connected systems. A security framework \cite{rahal2022antibotv} is proposed for efficient cyber-attack detection in the intravehicle networks (IVN) and External Vehicular Networks (EVNs). To differentiate cyberattack patterns more easily, Anbalagan et al.'s \cite{ anbalagan2023iids} IDS framework transforms vehicle network data into images, enhancing attack identification. However, both approaches rely on centralized learning, which raises privacy concerns within the broader context of the IoV, as data collection could potentially infringe on user privacy.

Uprety et al. ~\cite{ uprety2021privacy} introduced a FL-based collaborative IDS, enabling CAVs to train Deep Learning (DL) models on locally labeled datasets and share model parameters with a central FL server for global model aggregation. Similarly, Hbaieb et al. \cite{hbaieb2022federated} proposed an IDS for CAVs combining Software-Defined Networking (SDN) and FL, where SDN controllers train local models using data from CAVs, and global model aggregation is performed on a central cloud server. Both approaches ~\cite{ uprety2021privacy, hbaieb2022federated } rely on a central server, presenting risks as potential attack targets and single points of failure. There is also the risk of attackers intercepting and observing model updates during network transmission. Differently, Boualouache et al. \cite{boualouache2022federated} proposed an FL-based privacy-preserving IDS using multiple FL servers for global model aggregation. However, this solution, along with the previous ones ~\cite{uprety2021privacy, hbaieb2022federated}, do not consider the risk of a compromised FL client poisoning the model during training. To address these challenges, Zero-X employs differential privacy and blockchain technology, facilitating decentralized and privacy-enhanced federated training of the framework. Additionally, it incorporates a Byzantine Fault Tolerance consensus mechanism, named Proof-of-Accuracy (PoA), to ensure secure dissemination and aggregation of FL model updates.

Liu et al. \cite{liu2021blockchain} introduced an FL-based IDS that incorporates a blockchain-based incentive mechanism to mitigate adversarial attacks. In this system, CAVs function as FL clients, building models from their locally labeled datasets, with Roadside Units (RSUs) aggregating the global models. However, this approach's reliance on a simple MLP (Multi-Layer Perceptron) network for the classification task in intrusion detection may not be sufficiently effective in detecting new attack patterns. To classify various attack types, Abdel-Basset et al. \cite{abdel2021federated} presented FED-IDS, a Blockchain-enabled FL IDS utilizing a transformer network, capable of understanding the spatial and temporal patterns of traffic flows in vehicles. However, the authors acknowledge practical challenges with their consensus mechanism, such as divergence, limited efficiency, and high computational demands. Lai et al. \cite{lai2023improved} introduced a FL and Edge Cloud communication architecture (FL-EC) and a Feature Select Transformer (FSFormer) model for robust intrusion detection in the IoV. Yet, they recognize that this model comes with the drawback of high computational costs.

The aforementioned FL-based IDSs operate under the assumption that CAVs possess labeled datasets for FL rounds, an assumption that might be unrealistic due to the general absence of pre-labeled malicious traffic data. Additionally, these systems often assume an IID (Independent and Identically Distributed) configuration for classifier training, which may not reflect the significant imbalances in real-world training data, a known factor adversely affecting collaborative detection performance. Moreover, their dependence on closed-set and supervised learning methods limits the detection of unseen and zero-day attacks. In contrast, our framework employs a more realistic configuration, recognizing that CAVs typically do not have traffic labeling functions and considering the case of non-IID training data. Additionally, Zero-X enhances attack detection, including zero-day attacks, by integrating deep neural networks with Open-Set Recognition (OSR) capabilities.

To detect zero-day attacks in the IoV, Khan et al. \cite{khan2022enhanced} introduced an IDS using state-based Bloom filters and a bidirectional LSTM classifier for detecting cyber-attacks in both In-Vehicle and external networks. This approach, however, necessitates a considerable number of observations for high detection accuracy, and its detection rate for certain attack types is lower compared to other methods. Other recent studies \cite{jeong2023x, agrawal2022novelads, yang2021mth} have focused on detecting zero-day attacks in Intra-Vehicular Networks. Jeong et al. \cite{jeong2023x} utilized autoencoders to process live streams in CAN (Controller Area Network) message payloads. However, X-CANIDS \cite{jeong2023x} might face limitations in effectively detecting suspension attacks, which are typically identified by time-interval or sequence-based IDSs. In their respective works, Agrawal et al. \cite{agrawal2022novelads} and Yang et al. \cite{yang2021mth} employed unsupervised learning and anomaly-based IDS for CAN message analysis. While the approaches \cite{khan2022enhanced, jeong2023x, agrawal2022novelads} enable zero-day attack detection, they fall short in identifying the types of N-day attacks. Their effectiveness is further constrained by the lack of shared CAN databases among carmakers. Moreover, their dependence on centralized learning poses privacy concerns, where data collection could infringe on user privacy. In contrast, Zero-X, with its OSR detection model, enables the detection of zero-day attacks and the identification of N-day attacks. Additionally, it respects user privacy through federated learning, eliminating the need for data collection.
\color{black}

%To summarize, the main functionalities and components of the Zero-X framework proposed in this paper are listed in Table \ref{tab:sota}. Additionally, Table \ref{tab:sota} presents a clear comparison between 10 existing literature and the Zero-X framework based on the essential functionalities required for an effective and efficient vehicular network IDS.

\begin{figure*}[ht!]
    \centering
    \includegraphics[scale=0.35]{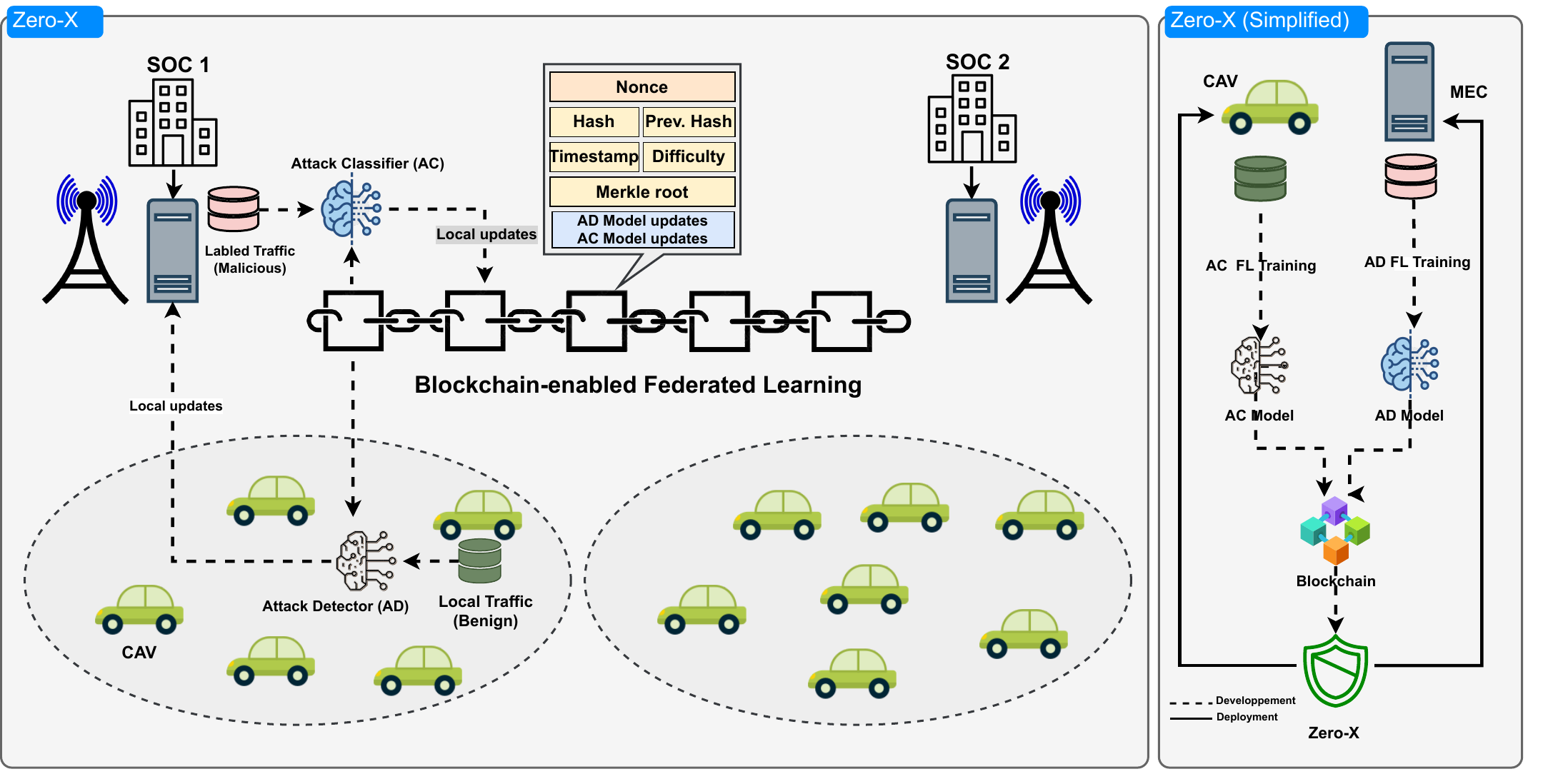}
    \caption{System design for the Zero-X framework}
    \label{fig:sd}
\end{figure*}

\section{System Design}\label{SD}
This section introduces the system model, seamlessly putting in collaboration different CAVs with SOCs. It illustrates the synergy operation between them through the Blockchain-enabled collaborative training  of the Zero-X Framework. Furthermore, an exploration of the system's addressed threat model will also be presented.

\subsection{System Model}
Our framework enables privacy-preserving collaboration among multiple CAVs and Security Operation Centers (SOC) that may be owned by different stakeholders within the ITS ecosystem, as illustrated in figure \ref{fig:sd}. To participate in the federated training process, the SOCs deploy MEC servers on strategically placed Base Stations (BSs). The system model is composed of three main elements with different roles which can be summarized as follows:

\begin{enumerate}[wide,  labelindent=10pt]
\item \textit{CAVs}: are tasked with training the Attack detector (AD) model using their own local dataset, which exclusively consists of benign traffic, as CAVs typically lack traffic labeling functions. When a CAV drives near a BS, it can download and read the last block of the blockchain to get the latest global AD model. Acting as FL workers, each CAV conducts the learning process using its own On Board Unit. The CAV trains the global model on its local data and adds a specific amount of Gaussian noise to the trained parameters before sending them back to the BS. This technique, known as differential privacy, protects the privacy of CAV's native data from honest-but-curious MEC nodes.  

\item \textit{MECs}: play a crucial role in validating and aggregating AD model updates that are received from CAVs. During the training of the Attack Classifier (AC) model, each MEC is assigned one of three roles: worker, validator, or miner. As a worker, the MEC trains the AC model using its local dataset, which includes labeled malicious network flows. Given that SOCs are equipped with cyber threat intelligence tools, it is reasonable to assume that they possess labeled samples of malicious network traffic. As a validator, the MEC ensures the accuracy of transactions and validates blocks that are proposed by miners. Finally, as a miner, the MEC is responsible for creating new blocks during the consensus process.

\item \textit{Blockchain}: to ensure secure and transparent FL model updates, the framework uses blockchain technology for AD and AC model updates sharing and aggregation (see figure \ref{fig:sd}). By establishing a consortium blockchain, SOCs can create a trusted and immutable ledger to store and securely distribute the AD and AC model updates. The decentralized nature of the blockchain ensures that no single SOC controls the entire network, thereby minimizing the risk of a single point of failure or data tampering.
\end{enumerate}

\subsection{Threat Model} \label{sec:threat}
We consider three categories of potential cybersecurity attacks that could target the IoV ecosystem. These include inter-vehicle attacks, attacks against the MEC infrastructure, and attacks against the intrusion detection mechanism itself.

\begin{itemize}
    \item \textit{Inter-vehicular attacks}: our threat model considers scenarios where one or colluding internal/external malicious CAVs launch DoS/DDoS attacks affecting the availability of the CAV and the network. For example, a colluding malicious CAVs can perform flooding attacks against a target CAV, aiming to overwhelming its capacity and preventing legitimate traffic from being processed. For comprehensive information on examples of attack scenarios please refer to \cite{boualouache2022survey}.
    
    \item \textit{Attacks against the MEC infrastructure}: MEC handles sensitive and valuable data, making it a target for various network attacks from both internal and external sources \cite{9757866}. As a crucial component of the IoV ecosystem its security is essential for ensuring the overall security of the IoV. Hackers could potentially create an IoV botnet \cite{antibot} by compromising a group of CAVs and using them to launch DDoS attacks on the MEC infrastructure. Such attacks are similar to DDoS attacks based on IoT bots.

    \item \textit{Attacks against the detection framework}: one potential threat against the security framework is attacks on the framework's detection capabilities. This could happen during the training phase, where malicious CAVs or compromised MECs may be selected as FL workers. If this occurs, the attackers could attempt to undermine the collaborative learning process by injecting malicious updates during the global model aggregation, which is known as a poisoning attack \cite{chen2021robust}. Additionally, attackers may also try to infer sensitive information from the model parameters through an inference attack \cite{zhao2022practical}. Both types of attacks can significantly compromise the accuracy and effectiveness of the security framework.
\end{itemize}

\begin{figure*}[ht!]
    \centering
    \includegraphics[scale=0.54]{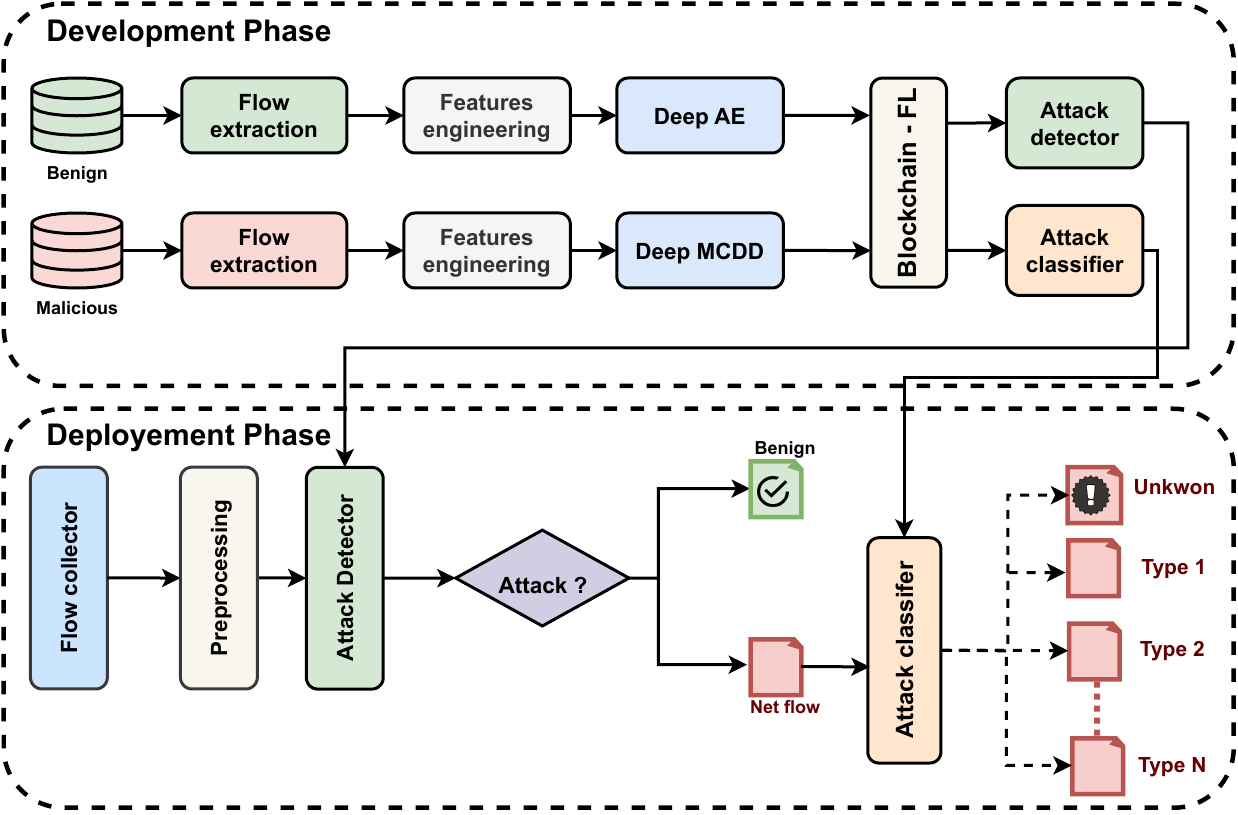}
    \caption{ Zero-X  framework workflow}
    \label{fig:archi}
\end{figure*}

\section{Proposed Zero-X framework}\label{SOL}
This section details the Zero-X framework. It begins with a presentation of the framework overview. Following this, the flow extraction and feature engineering processes are described. Then, the attack detection process and subsequent attack classification phase are explained. Finally, we will present the proposed blockchain-enabled FL scheme to ensure secure sharing and aggregation of model updates.

\subsection{Zero-X Overview}
During the development phase, the AD and AC models are trained independently on the network flows using a decentralized FL approach, as illustrated in Figure \ref{fig:archi}. First, network flows are extracted from collected raw network traffic, and then a set of features is calculated for each flow. The AD model is designed as a deep auto-encoder (DAE) trained exclusively on benign traffic data. On the other hand, the AC model is a deep multi-class data descriptor  \cite{deepmcdd} trained on the MEC's local dataset, which includes labeled malicious network flows. During the deployment phase,as illustrated in Figure~\ref{fig:archi}, the flow collector monitors incoming and outgoing packets. The preprocessing module extracts network flows from the raw traffic data, according to a specific Time Window (TW), and generates a vector of features. This features vector is then sent to the AD module, which evaluates the flow as benign or malicious. The framework's attack classifier module further examines any malicious flows detected by the AD to determine if they correspond to a new (0-day) or existing (N-day) attack. 
%In the following sections, we will outline the features engineering process. Subsequently, we will explain the attack detection process using a deep autoencoder. Next, we will elaborate on the attack identification process, which includes the use of a deep multi-class data descriptor to identify 0-day/N-day attacks. Finally, we will present a blockchain-enabled scheme to ensure secure sharing and aggregation of model updates during the FL process.

\subsection{Flow extraction \& features engineering} \label{sec:fe}
First, we use a combination of five properties from the packet header, including the network and transport layer headers of the TCP/IP protocol stack, to identify a traffic flow. These include the source IP address, the destination IP address, the source port number, and the destination port number, as well as the protocol. For each flow extracted, a set of features are calculated according to a given time window (ex. 10 seconds). Flow features include mainly packet header characteristics and statistics computed based on header information of network and transport layers. This set of features is then used as a features vector. 

\subsection{Attack Detector}

%% Add a figure of the AE stucture

%===== AutoEncoder
As a CAV runs a set of well-known applications such as those related to safety, convenience, and commercial use, its communication pattern should exhibit a high degree of regularity as long as it is not under attack or experiencing any faults. Conversely, any attack on the CAV's system would inevitably alter its communication pattern. Hence, we use a Deep Auto-Encoder (DAE) to model the expected communication pattern of the CAV and detect any attacks as anomalous occurrences. The DAE model \cite{ae} is an unsupervised model that compresses input vectors as code vectors using a set of recognition weights and then converts back to $m$ $(m < d)$ number of neurons reconstructed input vectors using a set of generative weights. An AE architecture has two major parts: the encoder and the decoder. The encoder reduces the dimension of the input vectors ($x_{i} \in R^{d}$) to numbers of neurons that form the hidden layer. The activation of the neuron $i$ in the hidden layer is given by: 

\begin{equation}
    \label{eq1}
h_{i}=f^{_{\theta }}(x)=s(\sum_{j=1}^{n}W_{ij}^{input}x_{j}+b_{i}^{input})
\end{equation}

where $x$ is the input vector, $\theta$ is the parameters 
$W^{input},b^{input}$, $W$  is an encoder weight matrix of dimension $m\times d$, while $b$ is a bias vector of dimension $m$. Thus, the input vector is encoded to a vector with fewer dimensions. The decoder maps the low-dimensional hidden representation $h_{i}$ to the original input space $R^{d}$  by the same transformation as the encoder. The function of mapping is as follows:
\begin{equation}
    \label{eq2}
x_{'}^{i}=g_{\theta' }(h) = s(\sum_{j=1}^{n}W_{ij}^{hidden}h_{j}+b_{i}^{hidden})
\end{equation}

The set of decoder parameters is $\theta'(W^{hidden}h_{j}+b^{hidden})$. The objective of an autoencoder is to minimize the reconstruction error relative to $\theta$ and $\theta'$ :

\begin{gather}
        \label{eq3}
\theta ^{*},\theta ^{'*}=arg_{\theta,\theta ^{'}} min \frac{1}{n}\sum_{i=1}^{n}\varepsilon (x_{i},x_{i}^{'}) \\
=arg_{\theta,\theta ^{'}} min \frac{1}{n}\sum_{i=1}^{n}\varepsilon (x_{i},g_{\theta ^{'}}(f_{\theta}(x_{i})))
\end{gather}

%===== AE for AD
The reconstruction error is utilized as the anomaly score. Network flows with significant reconstruction errors are regarded as malicious flows (anomalies). Only benign flows are used to train the DAE model. After training, the DAE model will reconstruct benign flows exceptionally well, but not malicious flows that it has never seen. Algorithm \ref{alg:algo_AE} shows the anomaly detection process using the reconstruction errors of the DAE model. The threshold, $\alpha$, is the mean squared error (MSE) median, and the sample's $C$ times the MSE Median Absolute Deviation (MAD) over the validation set. MAD uses the deviation from the median, which is less likely to be skewed by outlier values.

%========= Algorithm =========== 
\begin{algorithm}[h!]
\SetKwFunction{LocalUp}{\textbf{LocalUp}}
\SetKwFunction{ValUp}{\textbf{ValUp}}
\SetKwFunction{AggUp}{\textbf{AggrUp}}
\SetKwBlock{train}{Training phase:}{}
\SetKwBlock{test}{Testing phase:}{}
    \caption{Attack Detection}
    \label{alg:algo_AE}     
\train{
\KwIn{\textbf{$X_{Tr}$} : Train dataset, \textbf{$X_{V}$}: Validation dataset}
    $\phi, \theta \leftarrow$ train the DAE on $X_{Tr}$ \\
     \For{$i \in \{1,...,N\}$}{
         $RE_{V}[i]=\left \| x_{V}^{(i)}-g_{\theta }(f_{\phi}(x_{V}^{(i)})) \right \|$  
                } 
          $MAD = \text{median}(|RE_{V}[i]- \text{median}({MSE}_{RE_{V}})|)$ \\
     $\alpha = \text{median}({MSE}_{RE_{V}})) + \mathcal{C} \times MAD$

}\test{
    \KwIn{\textbf{$X_{Te}$}: Test dataset, \textbf{$\alpha$}: Threshold}
    %Test dataset $x^{(i)}$ $i \in \{1,...,N\}$, threshold $\alpha$\\
    
    %% replace by loss function 
    \For{$i \in \{1,...,N\}$}{
         $RE[i]=\left \| x_{Te}^{(i)}-g_{\theta }(f_{\phi}(x_{Te}^{(i)})) \right \|$  \\
         \eIf{$RE[i]> \alpha$}{$x_{Te}^{(i)}$ is a malicious flow \\
         Send $DP(x_{Te}^{(i)})$ to AC\\
         }{$x_{Te}^{(i)}$ is a benign flow \\
         Store $x_{Te}^{(i)}$ in $D_{k}$ for future retraining}
                }}
\end{algorithm}

% Complexity 
In Algorithm~\ref{alg:algo_AE}, the DAE training complexity is \( O(n \times m) \) (lines [1 - 2]), where \( n \) is the number of samples and \( m \) the count of weighted connections. RE computations for datasets (lines [3 - 4]) and median determination using a specialized algorithm (lines [5 - 6]) have a linear complexity of \( O(N) \). The testing phase functions \( g_0 \) and \( f_{\phi} \) (lines [7 - 9]) also operate in \( O(N) \), independent of the dataset size, with \( N \) being the test sample size.

\color{black}

\subsection{Attack Classifier}
%\subsubsection{Multi-Class Data Description}
%DeepMCDD Brief description 
To identify the type of the detected attack (N-day), or eventually determine if it is an unseen or 0-day attack, we leverage the potential of OSR through FL. In this paper, we use deep multi-class
data description method, called Deep-MCDD \cite{deepmcdd}, as it can accurately classify in-distribution (ID) samples into known classes and detect out-of-distribution (OOD) samples as well. This method seeks to identify, for each class, a spherical decision boundary that determines whether a test sample belongs to the class or not. Deep-MCDD integrates $K$ hypersphere-modeled one-class classifiers into a single network. The objective of learning a single hypersphere is to assess whether or not a test sample belongs to the target class; hence, training it for each class is useful for recognizing OOD samples that do not belong to any given class. Each sample from a \textit{K-th} class-conditional distribution can be considered as an isotropic Gaussian distribution with class mean $\mu_{k}$ and standard deviation $\sigma_{k}$ within the latent space, denoted by $\mathcal{N}(\mu_{k},\sigma_{k}^{2} I  )$. Using the resulting distributions, the class-conditional probabilities are calculated. This reflects the likelihood that an input sample is drawn from each distribution, and this probability can serve as a good confidence measure \cite{deepmcdd}. Based on the \textit{K-th} class conditional distribution, the distance function $D$ can be defined in Equation~\ref{eq:mcdd21}:

\begin{equation}
\label{eq:mcdd21}
\begin{aligned}
D_{k}(x) &\approx \frac{\left \|f(x;W)-\mu_{k}   \right \|^{2}}{2\sigma _{k}^{2}}+log\,  \sigma_{k}^{d}
\end{aligned}
\end{equation}

The Deep MCDD loss is defined as a Maximum A Posteriori (MAP) loss estimation of the generative classifier as defined in Equation~\ref{eq:mcdd22}:
\begin{dmath}
    \label{eq:mcdd22}
    Loss = \, \frac{1}{N}\sum_{i=1}^{N} \, log  \, \frac{exp(-D_{y_{i}} (x_{i})+b_{y_{i}})}{\sum_{k=1}^{k} exp(-D_{k} (x_{i})+b_{k})} 
\end{dmath}

The objective can be defined as follow:

\begin{dmath}
    \label{eq:mcdd3}
    \underset{W,\mu,\sigma,b}{min} \, \frac{1}{N}\sum_{i=1}^{N}\left [ D_{y_{i}} (x_{i}) \, -\, \frac{1}{\nu} \, log  \, \frac{exp(-D_{y_{i}} (x_{i})+b_{y_{i}})}{\sum_{k=1}^{k} exp(-D_{k} (x_{i})+b_{k})} \right ]
\end{dmath}
  
 The training can be seen as minimizing the intra-class deviation in the embedded space. The four trainable parameters used in the objective, that is, DNN weights ($W$), class means ($\mu_{class_{x}}$), standard deviations ($\sigma_{class_{x}}$), and biases ($b_{class_{x}}$), can be optimized simultaneously and effectively using minibatch SGD and gradient back-propagation. The confidence score $S(x)$ represents the distance between a test sample and the closest class-conditional distribution in the latent space. The  $C_{S}$ represents the minimum value of the confidence score calculated on the training set. If the confidence score $S(x)$  of a network flow $x$ is less than $C_{S}$, then it is considered  an unseen/ 0-day attack. Otherwise, we predict the class label of an input sample to the class with the highest posterior probability $\hat{y}(x)$. The procedure of training and testing the attack classifier is described in detail in Algorithm~\ref{alg:algo_AC}. 

The computational complexity of the training phase in Algorithm~\ref{alg:algo_AC} (lines [1 - 7]) for a dataset with \( N \) samples, \( k \) classes, and a latent space dimensionality of \( d \) is \( O(N \times k \times d) \). This includes squared difference and variance normalization operations per sample and class, as well as the cost of propagating samples through a neural network with \( m \) total connections, adding an extra complexity of \( O(N \times m) \). Therefore, the combined complexity for the dataset stands at \( O(N \times (m + k \times d)) \), prior to considering any potential optimizations from parallel processing and GPU acceleration that can be available in MEC infrastructure, since this model is trained by MECs. In the testing phase (lines [9 - 15]), each test sample is evaluated with a complexity of \( O(k \times d) \) due to distance computations for \( k \) classes in a \( d \)-dimensional feature space. The subsequent classification involves an arg max operation across classes with a complexity of \( O(k) \). The total complexity for testing \( N \) samples, therefore, is \( O(N \times k \times d) \).

%%% Attack identification 

%========= Algorithm =========== 

\begin{algorithm}[h!]
\SetKwBlock{trainId}{Training phase:}{}
\SetKwBlock{testId}{Testing phase:}{}
    \caption{Attack Classification}
    \label{alg:algo_AC}    
\trainId{
    \KwIn{ $X_{Tr}$ : Train dataset} 
    \textbf{Variables}: $k:$ number of attack classes, $W:$ Weights \\
    $f(.;W), \mu_{k}, \sigma_{k}, b_{k}  \leftarrow$ Train MCDD on $X_{Tr}$ \\
       Computing the confidence score $C_{S}$: \\
    \For{$x_{Tr}^{(i)} \in X_{Tr}$}{
    $S[x_{Tr}^{(i)}] = \underset{k}{min} D_{k}(x_{Tr}^{(i)})$
                } 
    $C_{S} = min(S)$
    }
\testId{
\KwIn{ $X_{Te}$: Test dataset, $C_{S}$}
\For{$x \in X_{Te}$}{
   $S(x) = \underset{k}{min} \, D_{k}(x)$
               
\eIf{$S(x) > C_{S}$}{
$\hat{y}(x) = arg\, \underset{k}{max} [-D_{k}(x)+b_{k}]$ \\
Find the $K^{th}$ closet type based on $\hat{y}(x)$ \\
}{

$x$ is a 0-day / Unseen attack
}
    } }    

\end{algorithm}

\subsection{Blockchain-enabled Federated Training} \label{sec:bc}
To ensure the secure and decentralized federated training of both AD and AC models, we use blockchain technology. In our framework, updates of both models are recorded on the same blockchain, while their training is performed independently. Integrating blockchain with FL requires careful selection of the blockchain type and consensus protocol, as they significantly impact scalability, latency, complexity, and cost. Public blockchains face issues like scalability limitations and high latency. Private blockchains improve scalability and latency but may reduce decentralization and privacy. Consortium blockchains, which Zero-X employs, strike a balance, making them well-suited for our system. Zero-X utilizes a consortium blockchain featuring a novel Byzantine Fault Tolerance consensus mechanism known as Proof-of-Accuracy (PoA). This mechanism ensures secure sharing and aggregation of model updates. Our approach strategically prioritizes scalability, latency, and security, with a slight compromise in decentralization. We consider two types of nodes within the blockchain network:

\begin{itemize}
  \item \textit{Passive node:} refers to CAVs that solely download and read blocks without actively participating in the consensus process.
  \item \textit{Active node:} refers to MECs actively participating in the consensus process.  In each round, every MEC is assigned to one of the following roles: worker, validator, and miner. A worker participates in the FL of the AC model and aggregates local AD model updates received from CAVs before relaying them. A validator node checks the correctness of transactions, and validates miner-proposed blocks. The miner is responsible for creating a new block during the consensus round.
\end{itemize}

The node's identifier (ID) is its public key. This key is used to verify the signatures of transactions or blocks created by the device. The following  outlines the process of selecting miners and validators in each round. The role-switching policy guarantees that miners and validators are re-selected for each new round, thereby minimizing the likelihood of a compromised MEC being repeatedly assigned to a validator or a miner role.

\subsubsection{Federated Training}
In our framework, updates of both models are recorded on the same blockchain. However, it is worth noting that sharing AD model updates from CAVs to MEC is done off-chain to enhance the scalability of the system. This approach allows faster and more efficient communication without overburdening the main blockchain with every AD update. We assume that a reference dataset $B_{Test}$ containing only benign network flows has been provided to all participating MECs. Each CAV extracts the AD model from the blockchain, then trains it on its local dataset. After local training, each CAV sends the model updates to the adjacent MEC node. Since an adversary may attempt to infer the data at the $i$-CAV from its uploaded updates, we use differential privacy to enforce privacy-preserving of local updates. After completing the local training, a $CAV_{i}$ will add a specific quantity of Gaussian noise to the trained parameters $w_{i}^{t}$. As outlined in Algorithm \ref{alg:bc1}, after clipping the gradient $w^{t}$: $w^{t+1} \leftarrow w^{t+1}/ \,\max (1, \frac{ ||w^{t+1}||}{C})$, where $C$ denotes the gradient norm bound, an additive noise $n_{i}^{t}$ is added to $w^{t}$. Then, the noised updates $\widetilde{w}_{i}^{t}$ are sent to the adjacent MEC node. To avoid model poising, the MEC node will first check the AD model updates $AD_{Up}$ on $B_{Test}$. The model is considered to be valid when the gap between $l(w^{t})$ and $l(w^{t-1})$ locates within a certain range $\delta$. 

When a MEC node is designated as a worker, it leverages its local dataset to train the AC model retrieved from the blockchain. Upon completing the training, the worker generates two transactions: $T_{AD}$, which includes the AD model update (computed based on the aggregated updates received from the CAVs), and $T_{AC}$, which contains the AC model's local updates. Both transactions are signed with the worker's private key and transmitted to the validator nodes.
The procedure of federated training of the AD and AC models is described in detail in Algorithm~\ref{alg:bc1}. 

\subsubsection{Proof-of-Accuracy Consensus Mechanism} \label{sec:poa}
In our proposed blockchain framework, collaboration among MECs is achieved through a consensus process where model updates are shared, packaged into blocks, and validated by a group of validators. To ensure the consensus process is efficient and secure, we introduce a lightweight consensus mechanism called Proof-of-Accuracy (PoA), which is inspired by the Delegated Byzantine Fault Tolerance (dBFT) algorithm \cite{dbft}. The PoA mechanism involves selecting a set of validator nodes (delegates) based on their contributions to improving the model's accuracy. We assume a total of $3f + 1$ validators, where $f$ is the maximum number of validators that can be faulty. Among these validators, one is designated as the miner. The detailed design of the PoA mechanism is explained in the following three steps.
\paragraph{Collecting and verifying transaction}
When a validator receives a transaction, it first verifies its signature. Then, it creates a test model $M_{Test}$ with the update, which it evaluates on its local dataset (for AC transactions) or on the reference dataset $B_{Test}$ (for AD transactions). Next, it calculates the accuracy gain  \textit{ACC\_gain} of  a transaction using the equation \ref{eq:acc_gain}, with the $G_{j-1}$ representing the global model constructed in round $R_{j-1}$. Validators exchange \textit{ACC\_gain} values for all transactions with each other. The final score of \textit{ACC\_gain} value for a given transaction is the sum of the \textit{ACC\_gain} values calculated by all validators. At the end of the validation process, all validators will have the same list of transactions with their respective \textit{ACC\_gain} values.

\begin{equation}
\text{ACC\_gain} = \text{ACC}(M_{\text{Test}}) - \text{ACC}(G_{j-1})
\label{eq:acc_gain}
\end{equation}

\paragraph{Generating Blocks }
The MEC assigned with the miner's role is responsible for creating a new block. Once the FL process has expired, the miner collects $T_{AD}$ and $T_{AC}$ transactions, including their corresponding \textit{ACC\_gain}, from the transaction pool. These transactions are then combined to form a candidate block. The miner signs the block with its private key, before broadcasting it to the other validators in the network.

%========= Algorithm =========== 
\begin{algorithm}[h!]
\SetKwBlock{DoParallel}{Do in parallel}{end}
\SetKwBlock{AdProc}{AD validation and aggregation:}{}
\SetKwBlock{AcProc}{AC Training:}{}
\SetKwBlock{genTrans}{Generating Transactions:}{}
\SetKwBlock{rcvTrans}{Processing received Transactions:}{}

\SetKwBlock{CAV}{CAV executes:}{}
\SetKwBlock{MEC}{MEC executes in parallel:}{\,}
\SetKwBlock{CS}{\CS executes in parallel:}{\,}
\SetKwFunction{Fun}{return}
\SetKwFunction{LocalUp}{\textbf{LocalUp}}
\SetKwFunction{ValUp}{\textbf{ValUp}}
\SetKwFunction{AggUp}{\textbf{AggrUp}}
\SetKwProg{FnLocal}{}{:}{\KwRet $w^{t+1}$ }
\SetKwProg{FnVal}{}{:}{\KwRet $Model_{VAL}$}
\SetKwProg{FnAgg}{}{:}{\KwRet $w^{t+1}$}

    \caption{Federated Training}
    \label{alg:bc1}
    \CAV{Collect benign $NF$ in dataset $L_{k}$ \\
    $\widetilde{AD}_{k}^{t+1} \leftarrow$ \textbf{LocalUp}($AD^{t}$,$E$, $\eta$, $L_{K}$, $\Delta_{\text{DP}} = True$) \\    
    send $\widetilde{AD}_{k}^{t+1}$ to adjacent MEC node}
    
    \MEC{
           \AdProc{Collect $AD$ updates within $\tau$ in $AD_{in}$  \\
      $VAL_{AD}^{MEC} \leftarrow$ \textbf{ValUp}($AD_{in}$, $\delta$, $T_{AD}$, $\Delta_{\text{DP}}= True$) \\
      $AD_{x}^{t+1} \leftarrow$ \textbf{AggUp}($AD^{t}, AD_{k}^{t+1}$) \\
        Create a transaction:  $T_{AD} \{AD_{x}^{t+1} || Sign_{M_{x}}\}$ \\
        Send $T_{AD}$ to Validators
      }

      \AcProc{
      Collect malicious $NF$ in dataset $M_{x}$ \\ 
      $AC_{x}^{t+1} \leftarrow$ \textbf{LocalUp}($AC^{t}$,$E$, $\eta$, $M_{x}$, $\Delta_{\text{DP}} = False$) \\  
        Create a transaction:  $T_{AC} \{AC_{x}^{t+1} || Sign_{M_{x}}\}$ \\
        Send $T_{AC}$ to Validators
    
      }
}

\FnLocal{\LocalUp{$w^{t}$, $E$, $\eta_{l}$, $D_{k}$, $\Delta_{\text{DP}}$}}{
$B \leftarrow$ split $\mathit{D_{k}}$ into batch of size $\mathit{B}$  \\
                \For{each local epoch $i \in \{1,...,E\}$}{
                 \For{batch $b$ $\in$ \textit{B}}{
                 $w^{t+1} \leftarrow w^{t} - \eta_{l} \bigtriangledown \mathit{\mathfrak{\mathit{l_{k}}}}
                 (w^{t},\mathit{D}_{k})$ \\
                 \textit{Clip the local updates}:
                 $w^{t+1} \leftarrow w^{t+1}/ \,\max (1, \frac{ ||w^{t+1}||}{C})$ \\
                \If{$\Delta_{\text{DP}}= True$}{
                 \textit{Add noise} \,
                 $\widetilde{w}^{t+1} \leftarrow w^{t+1} + n_{i}^{(t)}$
                 }
                }
                }

  }

    \FnVal{\ValUp{$w^{t}$, $\delta$, $D_{test}$}}{
    \If{$|l(w^{t})-l(w^{t-1})| \leq \delta$}{Add $w^{t}$ to $Model_{VAL}$}
        %\KwRet $Model_{VAL}$\;
    }

  \FnAgg{\AggUp{$w^{t}, w_{k}^{t+1}$}}{
   $w^{t+1}=\sum_{k=1}^{K} \frac{N_{K}}{N}\, w_{k}^{t+1}$ 
   }
\end{algorithm}

\paragraph{Consensus process}
Each validator examines the content of the candidate block, comparing the transactions and \textit{ACC\_gain} values with their own calculated values. If they match, the candidate block is accepted, and the validator signs the block with its private key as an endorsement. The endorsed candidate block is then sent to the miner. Once $2/3$ validators have signed the candidate block, the miner broadcasts the block along with the collected signatures to the other validators. Upon receiving the block from the miner, a validator verifies the signatures and checks if the block has enough signatures from validators to meet the required consensus threshold (2/3 validators). If the threshold is met, the block is considered final and is added to the blockchain. However, if the threshold is not met, the block is discarded, and the consensus process starts anew with a new round and a new miner. Once the block is successfully added to the blockchain, the validators propagate the block to other nodes in the network, and the nodes update their local copy of the blockchain to reflect the new block. Adding a new block triggers a new round of federated training of the two models (AD and AC). Model training stops when the average value of the model's \textit{ACC\_gain} (calculated from the last block) approaches 0. The top \textit{3f+1} worker nodes with the highest \textit{ACC_gain} in the current round will be selected as validators for the next round. On the other hand, nodes that have accumulated a negative \textit{ACC_gain} over a specified number of rounds will be considered unreliable and excluded from participating in subsequent training and consensus rounds. This measure helps eliminate poor updates that could hinder the convergence of the global models. Importantly, it also protect the training process against potential model poisoning attacks that may arise from a compromised MEC.

\subsubsection{Incentive/ reward mechanisms}
The SOCs participating in the federated training are intrinsically motivated by the advantage of having an updated IDS, which obviates the need for supplementary incentives or reward mechanisms. However, to promote the active engagement of CAVs in the learning process, it is vital to establish an incentive mechanism. In this regard, we have adopted the Stackelberg game theoretic-based incentive mechanism proposed in our previous research \cite{boualouache2022federated}.
%%% 

\section{Performance Evaluation}\label{SIM}

% \usepackage{booktabs}
In this section, we will assess the effectiveness of the Zero-X framework by conducting a thorough evaluation on two recent datasets. The first dataset, 5G-NIDD\cite{5GNIDD}, consists of 5G network traffic traces of attacks that target the MEC infrastructure. The second dataset, VDoS \cite{VDoS}, contains network traffic originating from inter-vehicle attacks. After briefly introducing the datasets and the experimental setting, we analyze the detection performance in detail. We also test various time-window (TW) sizes as it directly impacts the detection delay. Further, we will comprehensively analyze the identification performance of N-day and 0-day attacks. We then evaluate the performance of the blockchain system, and present the incentive mechanism's performance. Next, we present a security analysis of the Zero-X framework. Finally, we compare our approach with existing solutions in the literature. 

\subsection{Datasets}
The 5G Network-Intrusion Detection and Defense (5G-NIDD) dataset \cite{5GNIDD} comprises data collected from a 5G testbed connected to the 5G Test Network at the University of Oulu in Finland. The testbed consists of two base stations, each equipped with an attacker node and multiple benign 5G users. The attacker nodes simulate various attack scenarios against the server deployed in the 5GTN Mobile Edge Computing (MEC) environment. The dataset includes examples of Denial of Service (DoS) attacks, such as ICMP Flood, UDP Flood, SYN Flood, HTTP Flood, and Slowrate DoS, as well as port scans, including SYN Scan, TCP Connect Scan, and UDP Scan. The VDoS dataset \cite{VDoS}, is a publicly available dataset that contains both benign and malicious network traffic data. The dataset was generated using a realistic testbed comprising two vehicles, physical and virtual machines, access points, and Cisco antennas. Three different scenarios, urban, rural, and highway, were used to collect the network traffic data. The dataset contains the network traces of three types of DoS attacks: UDP Flood, SYN Flood, and Slowloris. The sample distribution of network traffic per attack for both datasets is presented in Table~\ref{tab:dataset}.

% \usepackage{booktabs}
% \usepackage{multirow}

\subsection{Experimental setting}
% Flow extraction & features engineering
To extract flows and calculate features from raw traffic (PCAP files), we developed some scripts using CICFlowMeter \cite{cic}, a popular flow traffic exporters. For a more comprehensive understanding of the list of features and their corresponding descriptions, kindly refer to  \cite{korba2023federated}. We perform data preprocessing, which includes data cleaning, data normalization, and feature engineering. To ensure accurate results, we follow a specific protocol for splitting the datasets. Legitimate traffic is divided into three parts: 60\% for training the AD, 20\% for AD validation, and 20\% for testing. Similarly, malicious traffic is split into 60\% for training the AC, 20\% for AC validation, and 20\% for testing. The dataset splitting is illustrated in Table \ref{tab:split}. To create a realistic test scenario, we designate one type of attack as the 0-day attack, while the remaining attack types are classified as N-day attacks. The training set excludes samples from the 0-day attack, and the AC is trained on N-day attacks for classifying inputs into $K-1$ attack types. The test set comprises all samples from the 0-day attack and N-day attack samples reserved for testing. This process is repeated with different 0-day attacks, resulting in $K$ scenarios per dataset. 

We implemented a DAE with five hidden layers. The local models' weights were computed using the Stochastic Gradient Descent (SGD) algorithm. The Mean Squared Error (MSE) loss function was used, with a learning rate of 0.012. After each round, the weights of the global model were calculated using the Federated Averaging algorithm \cite{FedAvg}. We trained and tested the model in the Google Colab cloud environment. We used the Pytorch package to implement the local and federated learning models. 
\begin{table}[]
\centering
\color{black}
\caption{Dataset samples distribution}
\label{tab:dataset}
\begin{tabular}{@{}llll@{}}
\toprule
Dataset                  & Attack      & Absolute count & Fraction \\ \midrule
\multirow{9}{*}{5G-NIDD} & Benign      & 477737         & 39.29\%  \\
                         & UDPFlood    & 457340         & 37.61\%  \\
                         & HTTPFlood   & 140812         & 11.58\%  \\
                         & SlowrateDoS & 73124          & 6.01\%   \\
                         & TCPConScan  & 20052          & 1.65\%   \\
                         & SYNScan     & 20043          & 1.65\%   \\
                         & UDPScan     & 15906          & 1.31\%   \\
                         & SYNFlood    & 9721           & 0.80\%   \\
                         & ICMPFlood   & 1155           & 0.09\%   \\ \midrule
\multirow{4}{*}{VDoS}    & Benign      & 101615         & 42.06\%  \\
                         & UdpFlood    & 68953          & 28.54\%  \\
                         & SynFlood    & 62354          & 25.81\%  \\
                         & Slowloris   & 8689           & 3.60\%   \\ \bottomrule
\end{tabular}
\end{table}

% \usepackage{booktabs}
\begin{table}[]
\centering
\caption{Dataset splitting}
\label{tab:my-table}
\begin{tabular}{@{}llllll@{}}
\toprule
\textbf{Samples} & \textbf{Train  AD} & \textbf{Val AD} & \textbf{Train AC} & \textbf{Val AC} & \textbf{Test} \\ \midrule
Benign           & 60\%               & 20\%            & -                 & -               & 20\%          \\
N-day            & -                  & -               & 60\%              & 20\%            & 20\%          \\
0-day            & -                  & -               & -                 & -               & 100\%         \\ \bottomrule
\end{tabular}
\label{tab:split}
\end{table}

\subsection{Attack detection performance}
Firstly, we evaluate the accuracy of the global model AD on the two datasets, considering the IID and Non-IID configurations. Next, we conduct a thorough analysis of the model's performance using various metrics. Finally, we test multiple levels of differential privacy to determine the optimal balance between privacy and model performance. To obtain a comprehensive and insightful evaluation of attack detection performance, we consider the following metrics:
\begin{itemize}
\item \textit{Precision}: the ratio of classified malicious flows that are truly malicious:
            \begin{equation}
             Precision=\frac{TP}{TP+FP}
	        \label{eq:prec}
	       \end{equation}
	       
\item \textit{Accuracy}: the ratio of correctly predicted malicious samples to the total samples:
 	      \begin{equation}
            Accuracy =\frac{TP+TN}{TP+FN+FP+TN}
	        \label{}
	       \end{equation}

\item \textit{TPR}: measures the proportion of benign flows that are incorrectly classified as malicious by the model:
   \begin{equation}
             TPR=\frac{TP}{TP+FN}
	        \label{eq:Rec}
	       \end{equation}
\item \textit{FPR}: measures the effectiveness of the model in incorrectly recognizing normal flow as an anomaly:
    \begin{equation}
    FPR = \frac{FP}{FP + TN}
    \label{eq:fpr}
    \end{equation}

\item \textit{F1-Score}: is the harmonic average of the precision and recall:
            \begin{equation}
            F1\_Score = \frac{2*(TPR * Precision)}{(TPR + Precision)}
	        \label{eq:f1}
	       \end{equation}

\item \textit{AUCROC}: measures how effective the model is in distinguishing between normal and malicious flows: 
           \begin{equation}
 AUROC =  \int_{0}^{1} \text{TPR}(\text{FPR}) \, d\text{FPR}
  %  AUCROC = \frac{1}{2} \left(\frac{TP}{TP+FN}+ \frac{TN}{TN+FP} \right)
	        \label{eq:auc}
	       \end{equation}

\end{itemize}

 TP, TN, FP, and FN denote true positive, true negative,
false positive, and false negative, respectively. 

\begin{table}[h!]
  %\begin{center}
    \caption{Detection performance}
    \label{tab:tabDec}
    \begin{adjustbox}{width=0.49\textwidth}
    \begin{tabular}{l|S|r|r|r|r}
      \toprule
      \textbf{Dataset} & \textbf{Dist} & \textbf{Accuracy (\%)} & \textbf{Precision (\%)} & \textbf{TPR (\%)} & \textbf{F1-Score (\%)}\\
      \midrule
      \multirow{2}{*}{5G-NIDD} & {IID} & 92.27 & 95.51 & 93.05 & 94.26\\
      & {Non-IID} & 92.43 & 94.63 & 94.26 & 94.44\\
      \midrule
      \multirow{2}{*}{VDoS} & {IID} & 97.75 & 97.68 & 98.61 & 98.17\\
      & {Non-IID} & 97.61 & 97.07 & 97.75 & 97.63\\
      \bottomrule
    \end{tabular}
    \end{adjustbox}
 % \end{center}
\end{table}

\begin{figure*}[ht]
\centering
\begin{subfigure}[b]{0.41\textwidth}
\begin{adjustbox}{width=1\textwidth, center}
\includegraphics{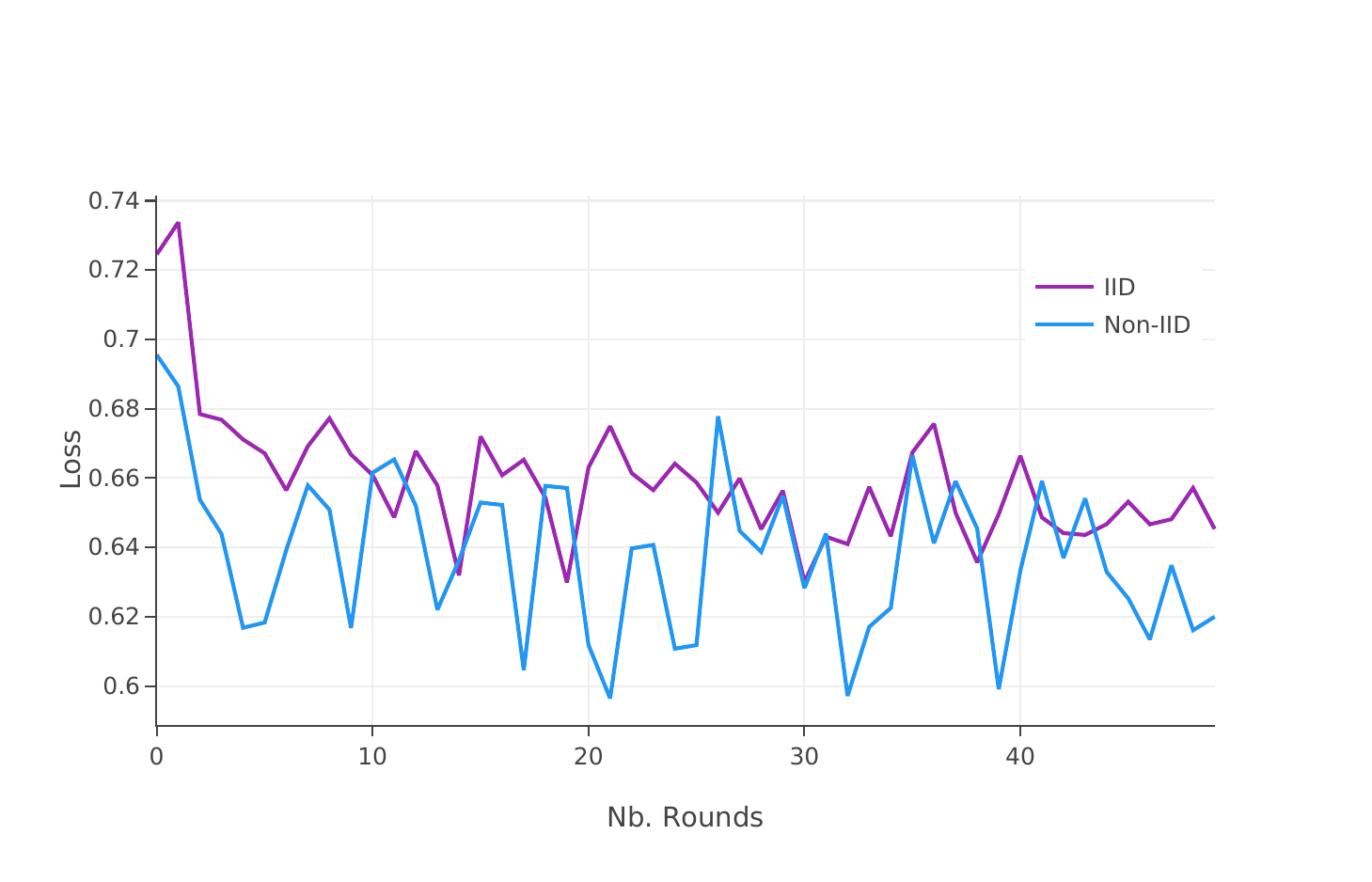}
\end{adjustbox}
\label{fig:lossAEVdoS}
\caption{VDoS - AD Loss}

\end{subfigure}
\begin{subfigure}[b]{0.41\textwidth}
\begin{adjustbox}{width=1\textwidth, center}
\includegraphics{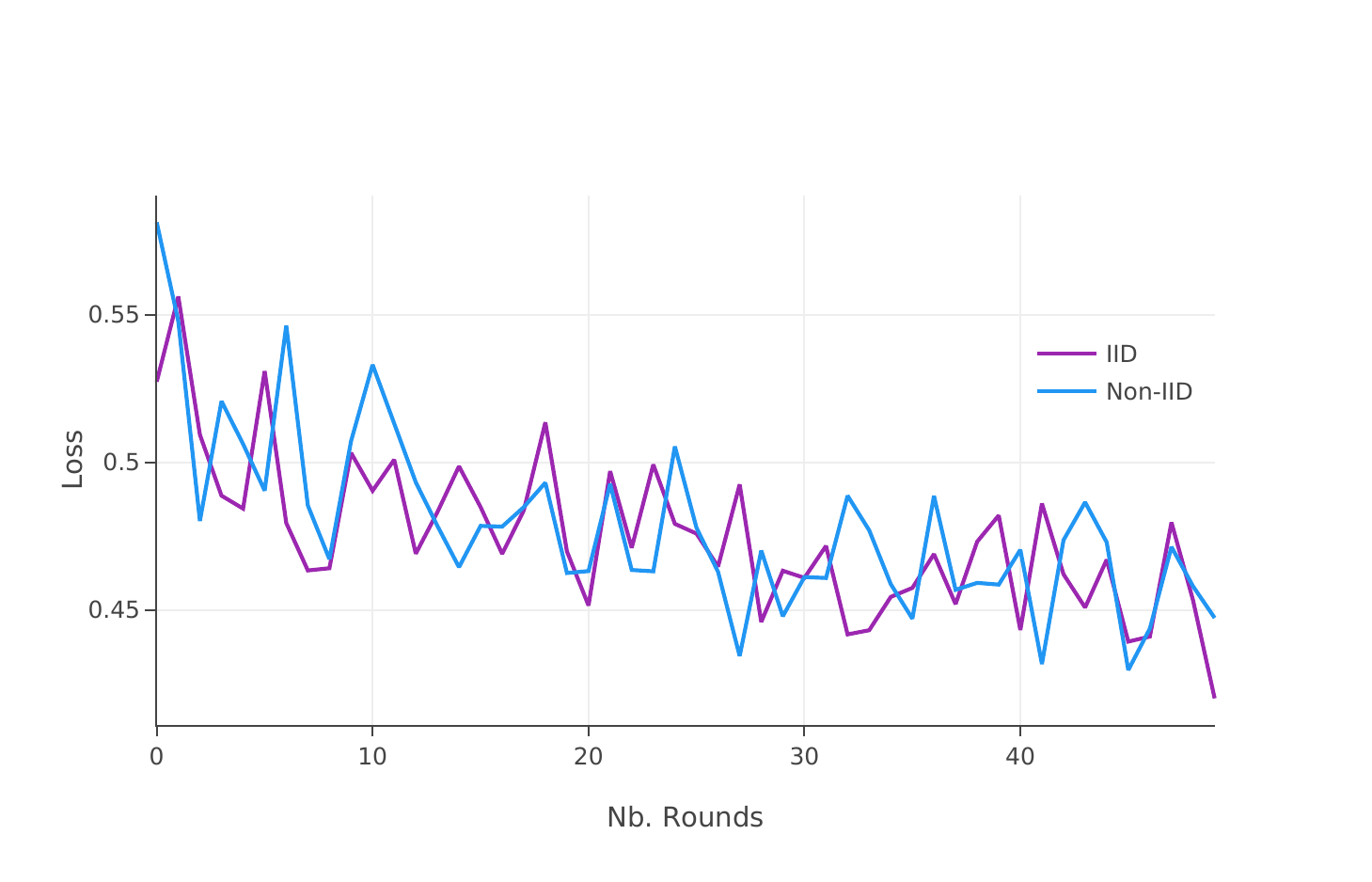}
\end{adjustbox}
\label{fig:lossAE5G}
\caption{5G-NIDD - AD Loss}
\end{subfigure}
\caption{Learning performance of AD models}
\label{fig:lossAD}
\end{figure*}

\begin{figure*}
\centering
\begin{subfigure}[b]{0.35\textwidth}
\includegraphics[width=\textwidth]{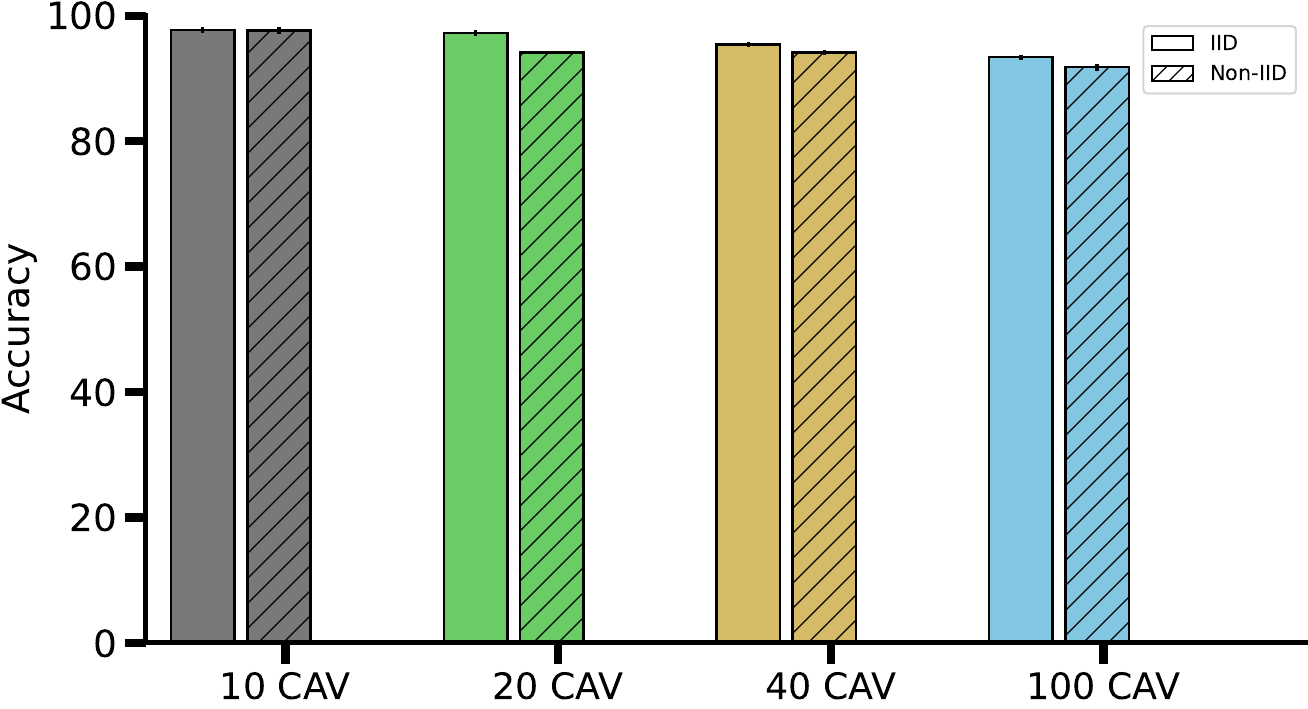}
\caption{Accuracy}
\end{subfigure}
\begin{subfigure}[b]{0.35\textwidth}
\includegraphics[width=\textwidth]{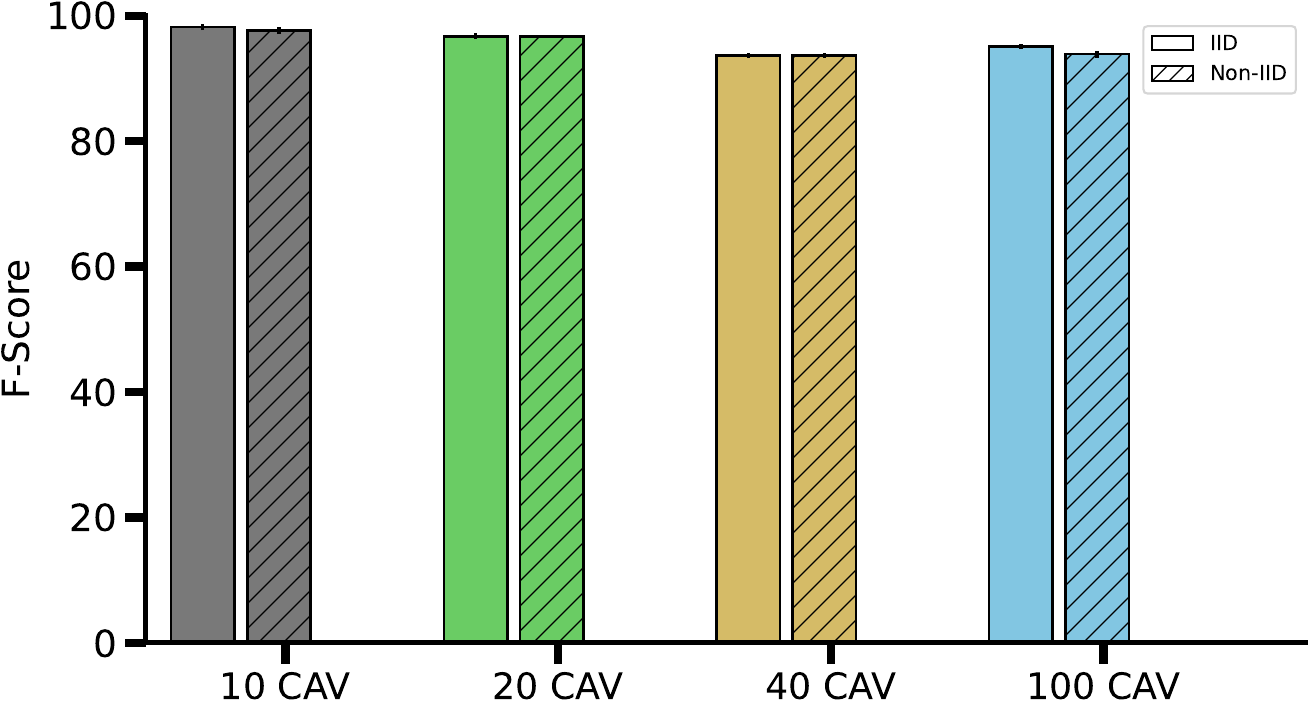}
\caption{F1-Score}
\end{subfigure}
\caption{AD performance versus the number of CAVs: Accuracy and F1-score analysis}
\label{fig:nbcav}
\end{figure*}

Figure \ref{fig:lossAD} illustrates the AD loss function over the number of federated learning rounds, which typically exhibits a decreasing trend as the model is trained with more data. Although slight fluctuations may be observed, the function stabilizes for both datasets and with both IID and Non-IID configurations. After the $10^{th}$ round, we observe minor improvements in the model's accuracy. The results as shown in Table \ref{tab:tabDec} are quite promising. They suggest that the model exhibits strong and consistent performance across both datasets and data distribution scenarios. The minor differences observed between the IID and Non-IID conditions indicate that the model is resilient and maintains its robustness, even when faced with different data distributions.

We tested different configurations by varying the number of CAVs per FL round. Figure \ref{fig:nbcav} illustrates that an increasing number of CAVs tends to slightly decrease the accuracy and F1-score, particularly for the Non-IID distribution. This could be due to the data becoming more diverse as more CAVs are involved, making it more challenging for the model to find a global optimum. As expected, the results consistently demonstrate that the IID distribution outperforms the Non-IID distribution in terms of accuracy and F1-score, regardless of the number of CAVs.

\begin{table*}[]
\centering
\caption{Detection performance with DP}
\begin{tabular}{@{}llllllll@{}}
\toprule
\textbf{Distribution}             & \textbf{DP}  & \textbf{Accuracy (\%)} & \textbf{Precision (\%)} & \textbf{TPR (\%)}     & \textbf{FPR (\%)}    & \textbf{F1-Score (\%)} & \textbf{AUROC (\%)}     \\ \midrule
\multirow{4}{*}{\textbf{IID}}     & No-DP            & \textbf{97.68}  & 97.75            & 98.61          & \textbf{3.92} & \textbf{98.17}  & \textbf{97.34} \\
                                  & ($\epsilon$ = 1.0)    & 97.55           & 97.37            & \textbf{98.80} & 4.61          & 98.08           & 97.10          \\
                                  & ($\epsilon$ = 0.1)  & 95.30           & 96.92            & 95.62          & 5.25          & 96.12           & 95.19          \\
                                  & ($\epsilon$ = 0.01) & 94.90           & 96.98            & 94.93          & 5.15          & 95.76           & 94.89          \\ \midrule
\multirow{4}{*}{\textbf{Non-IID}} & No-DP            & 97.07           & 97.61            & 97.75          & 4.11          & 97.63           & 96.82          \\
                                  & ($\epsilon$ = 1.0)    & 96.84           & \textbf{97.92}   & 97.07          & 3.57          & 97.42           & 96.75          \\
                                  & ($\epsilon$ = 0.1)  & 95.40           & 96.89            & 95.79          & 5.28          & 96.24           & 95.26          \\
                                  & ($\epsilon$ = 0.01) & 92.63           & 97.41            & 90.79          & 4.20          & 93.74           & 93.29          \\ \bottomrule
\end{tabular}
\label{tab:dp}
\end{table*}

According to the results presented in Table \ref{tab:dp}, the AD model exhibited a high detection rate and a low false positive rate. This indicates that the model effectively distinguished between benign and malicious network traffic in both configurations (IID and Non-IID). Table \ref{tab:dp} further demonstrates the impact of differential privacy level on the performance of the AD model. The differential privacy level is varied by changing the epsilon ($\epsilon$) value, which controls the amount of noise added to the model updates. As the differential privacy level increases (i.e., epsilon decreases), the precision, accuracy, TPR, F1-score, and ROC all tend to decrease. This is expected because increasing the amount of noise added to the model update will make it more difficult for the model to detect malicious traffic accurately. By setting the value of $\epsilon$ to 1, DP only slightly impacts system performance. For the IID configuration, we observe a degradation of 0.24\% in AUROC, while for the Non-IID configuration, we observe a degradation of 0.07\%. It is worth noting that even with the most challenging configuration - Non-IID with a high noise level (epsilon=0.01) - the model still achieves an accuracy higher than 92\% and a precision of 97\%. The fact that DP has a relatively small effect on the model's performance can be attributed to the fact that detection also depends on the threshold $\alpha$. Even if the addition of noise increases the loss of the model, the threshold will still be calculated based on the new loss and remain slightly affected.

% \usepackage{booktabs}
% \usepackage{multirow}

\subsection{Detection delay} \label{sec:tw}
\begin{figure}[]
    \centering
    \includegraphics[scale=0.35]{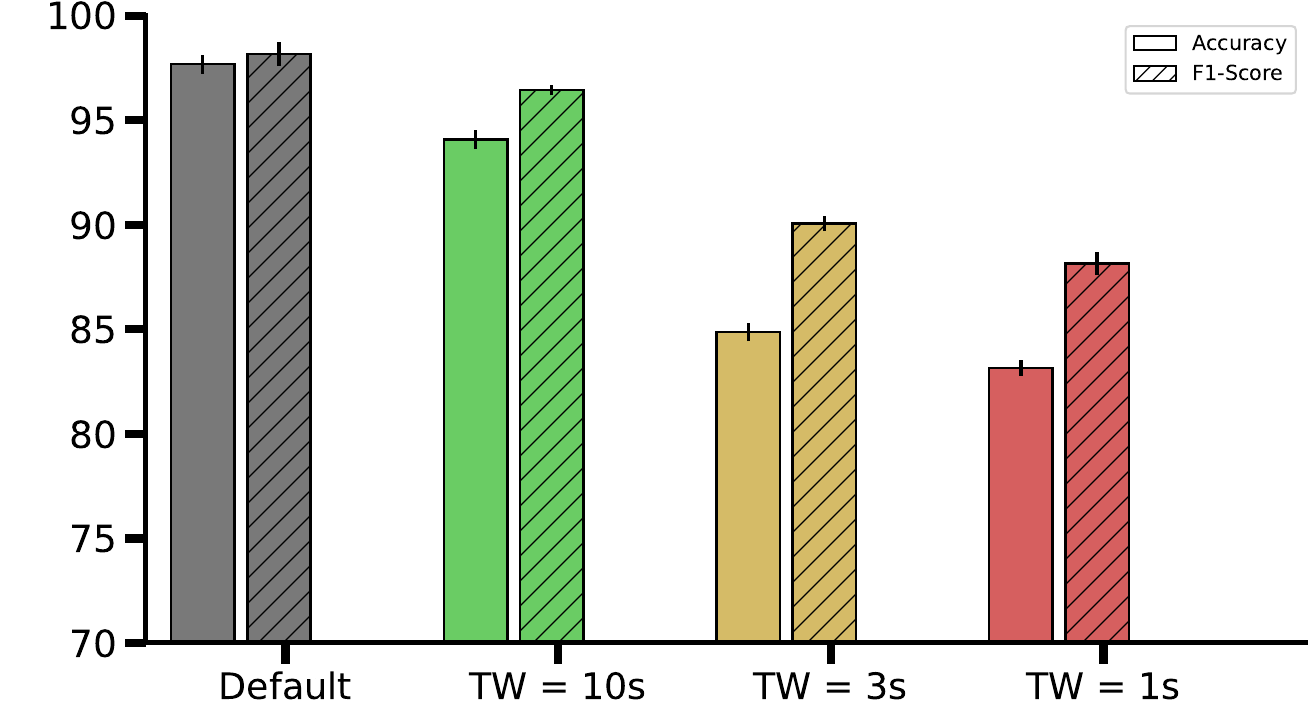}
    \caption{The impact of TW size on the detection accuracy}
    \label{fig:TW}
\end{figure}
The time-window (TW) size is a critical parameter that has a direct impact on detection delay. While a smaller TW can reduce detection delay, it may not capture certain traffic properties that require a longer time interval to become detectable. Therefore, we conducted experimental tests using multiple TW sizes to determine the optimal one.

The results presented in Figure \ref{fig:TW} reveal a correlation between the TW and the accuracy and F1-score of the AD model. Specifically, as the TW decreases, both performance metrics deteriorate. The default window size, which corresponds to the flow lifetime, yields the highest accuracy and F1-score. However, if a smaller window size is desired, the 10-second window remains the best option as it still maintains a high level of accuracy and F1-score. The 1-second window, on the other hand, performs the poorest among the tested windows, with an accuracy of 83.16\% and an F1-score of 88.15\%. Nevertheless, it still provides satisfactory detection capabilities. Ultimately, the choice of TW will depend on the desired level of security when deploying the framework, including considerations such as the tolerance for false negatives and the acceptable detection delay.

\begin{table*}[]
\color{black}
\centering
\caption{N-Day Attack identification}
\begin{tabular}{llS[table-format=3.2]S[table-format=3.2]S[table-format=3.2]S[table-format=3.2]S[table-format=1.2]S[table-format=1.2]S[table-format=3.2]S[table-format=3.2]}
\toprule
{\textbf{Dataset}} & {\textbf{Attack}} & \multicolumn{2}{c}{{\textbf{Precision (\%)}}} & \multicolumn{2}{c}{{\textbf{TPR (\%)}}} & \multicolumn{2}{c}{{\textbf{FPR (\%)}}} & \multicolumn{2}{c}{{\textbf{F1-Score (\%)}}} \\
\cmidrule(lr){3-4} \cmidrule(lr){5-6} \cmidrule(lr){7-8} \cmidrule(lr){9-10}
                   &                   & {IID} & {Non-IID} & {IID} & {Non-IID} & {IID} & {Non-IID} & {IID} & {Non-IID} \\
\midrule
\multirow{8}{*}{5G-NIDD} & HTTP Flood      & 100.00 & 99.93 & 100.00 & 100.00 & 0.00 & 0.02 & 100.00 & 99.96 \\
                         & UDP Flood       & 100.00 & 100.00 & 100.00 & 100.00 & 0.00 & 0.00 & 100.00 & 100.00 \\
                         & SYN Flood       & 99.50 & 100.00 & 100.00 & 100.00 & 0.01 & 0.00 & 99.74 & 100.00 \\
                         & ICMP Flood      & 100.00 & 100.00 & 100.00 & 100.00 & 0.00 & 0.00 & 100.00 & 100.00 \\
                         & SlowrateDoS     & 97.01 & 98.76 & 97.95 & 95.21 & 0.34 & 0.43 & 97.46 & 96.90 \\
                         & UDP Scan        & 99.69 & 99.41 & 100.00 & 100.00 & 0.01 & 0.02 & 99.84 & 99.70 \\
                         & TCP Con Scan& 99.74 & 92.24 & 99.21 & 99.22 & 0.01 & 0.26 & 99.46 & 95.49 \\
                         & SYN Scan        & 99.75 & 100.00 & 99.75 & 99.50 & 0.00 & 0.00 & 99.87 & 99.75 \\
\midrule
\multirow{3}{*}{VDoS}    & SYN Flood       & 99.99 & 100.00 & 100.00 & 100.00 & 0.06 & 0.00 & 100.00 & 100.00 \\
                         & Slowloris       & 100.00 & 100.00 & 99.94 & 99.94 & 0.00 & 0.00 & 99.97 & 99.97 \\
                         & UDP Flood       & 100.00 & 100.00 & 100.00 & 100.00 & 0.00 & 0.00 & 100.00 & 100.00 \\
\bottomrule
\end{tabular}
\label{tab:N-day}
\end{table*}

\subsection{Attack identification performance}
The framework's attack classifier module further examines any malicious flows detected by the AD to determine if they correspond to a new (0-day) or existing (N-day) attack. We evaluate the accuracy of the global AC model on the two datasets, considering both IID and non-IID configurations. Additionally, we conduct a comprehensive analysis of the model's performance in detecting both N-day and 0-day attacks. 

\subsubsection{N-day attack identification performances}
Overall, the AC model successfully detects various types of attacks, even when confronted with different data distributions, as illustrated in Table \ref{tab:N-day}. For the 5G-NIDD dataset, the AC model achieved high performance, with precision and TPR scores of 100\% in most cases. The FPR was also low, indicating that the model did not classify benign traffic as malicious. However, for SlowrateDoS and TCPConnectScan attacks with Non-IID data distribution, the F1-score was slightly lower compared to other cases. The slow nature of traffic makes it challenging to detect the Slowrate DoS attack, which differs from other DoS attacks in terms of speed and packet volume sent. In a TCPConnectScan attack, the three-way handshake process is successfully established, mimicking a legitimate connection and making the attack challenging to detect. However, it is worth noting that the detection rate for this type of attacks remains very high, with a detection rate of over 95\%. For the VDoS dataset, the AC model achieved a perfect score in all cases with both IID and Non-IID data distributions.

% \usepackage{multirow}
\begin{table*}[]
\caption{0-Day Attack identification}
\centering
\begin{tabular}{rlrrrr}
\hline
\textbf{Dataset}                   & \textbf{Attack}                 & \textbf{Distribution} & \textbf{TNR85 (\%)} & \textbf{AUROC (\%)} & \textbf{Accuracy (\%)} \\ \hline
\multirow{16}{*}{\textbf{5G-NIDD}} & \multirow{2}{*}{HTTPFlood}      & IID                   & 82.65        & 89.86        & 88.51        \\
                                   &                                 & Non-IID               & 79.84        & 90.91        & 90.92        \\ \cline{2-6} 
                                   & \multirow{2}{*}{UDPFlood}       & IID                   & 79.90        & 87.21        & 85.64        \\
                                   &                                 & Non-IID               & 66.56        & 76.32        & 77.50        \\ \cline{2-6} 
                                   & \multirow{2}{*}{SYNFlood}       & IID                   & 90.88        & 90.42        & 91.53        \\
                                   &                                 & Non-IID               & 96.21        & 95.49        & 93.30        \\ \cline{2-6} 
                                   & \multirow{2}{*}{ICMPFlood}      & IID                   & 100.00       & 100.00       & 100.00       \\
                                   &                                 & Non-IID               & 100.00       & 92.32        & 96.12        \\ \cline{2-6} 
                                   & \multirow{2}{*}{SlowrateDoS}    & IID                   & 1.67         & 11.57        & 50.62        \\
                                   &                                 & Non-IID               & 0.06         & 0.05         & 50.00        \\ \cline{2-6} 
                                   & \multirow{2}{*}{UDPScan}        & IID                   & 98.56        & 98.53        & 96.77        \\
                                   &                                 & N-IID               & 100.00       & 99.32        & 97.94        \\ \cline{2-6} 
                                   & \multirow{2}{*}{TCPConnectScan} & IID                   & 100.00       & 99.24        & 99.41        \\
                                   &                                 & Non-IID               & 99.95        & 96.23        & 98.03        \\ \cline{2-6} 
                                   & \multirow{2}{*}{SYNScan}        & IID                   & 99.95        & 98.51        & 97.19        \\
                                   &                                 & Non-IID               & 93.20        & 96.89        & 93.73        \\ \hline
\multirow{6}{*}{\textbf{VDoS}}     & \multirow{2}{*}{SYNFlood}       & IID                   & 99.18        & 99.31        & 98.65        \\
                                   &                                 & Non-IID               & 99.43        & 99.66        & 99.50        \\ \cline{2-6} 
                                   & \multirow{2}{*}{Slowloris}      & IID                   & 88.04        & 93.12        & 93.43        \\
                                   &                                 & Non-IID               & 87.29        & 90.67        & 93.48        \\ \cline{2-6} 
                                   & \multirow{2}{*}{UDPFlood}       & IID                   & 27.10        & 55.68        & 61.79        \\
                                   &                                 & Non-IID               & 44.51        & 71.39        & 71.61        \\ \hline
\end{tabular}
\label{tab:0-day}
\end{table*}

\subsubsection{0-day attack identification performances}
As previously mentioned, to create a realistic testing scenario, we designate one type of attack as the '0-day' attack, while classifying the remaining attack types as 'N-day' attacks. For each dataset, we repeat this process with different 0-day attacks, resulting in $K$ scenarios per dataset. To evaluate our model's performance, we use standard metrics such as accuracy and AUROC. Additionally, we use the true negative rate (TNR) at 85\% true positive rate (TPR), denoted as TNR85, a performance metric commonly used for out-of-distribution (OOD) detection \cite{deepmcdd}.

\begin{figure*}
\centering
\label{fig:bc}
\begin{subfigure}[b]{0.31\textwidth}
\begin{adjustbox}{width=1\textwidth, center}
\includegraphics{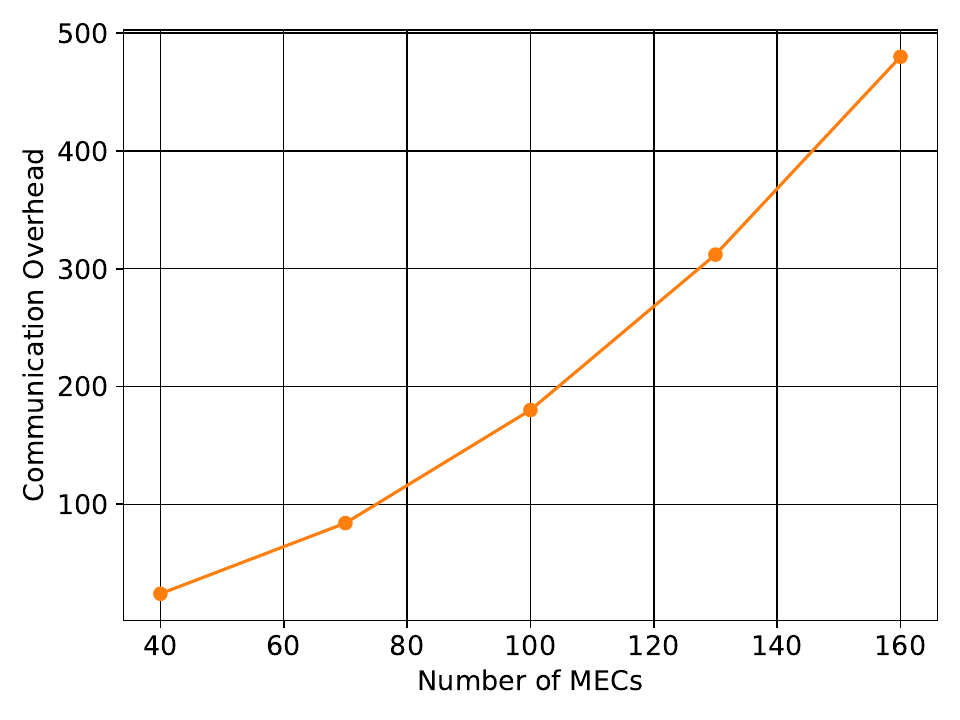}
\end{adjustbox}
\caption{Messages Overhead}
\label{fig:overhead}
\end{subfigure}
\begin{subfigure}[b]{0.3\textwidth}
\begin{adjustbox}{width=1\textwidth, center}
\includegraphics{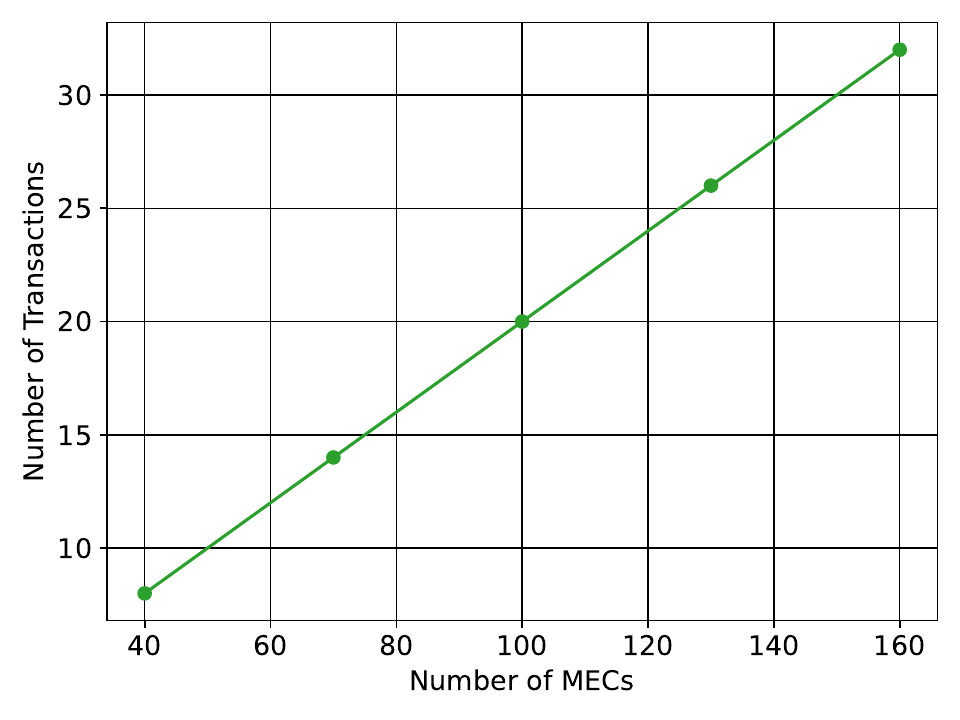}
\end{adjustbox}
\caption{Nb. Transactions per block}
\label{fig:trans}
\end{subfigure}
\begin{subfigure}[b]{0.3\textwidth}
\begin{adjustbox}{width=1\textwidth, center}
\includegraphics{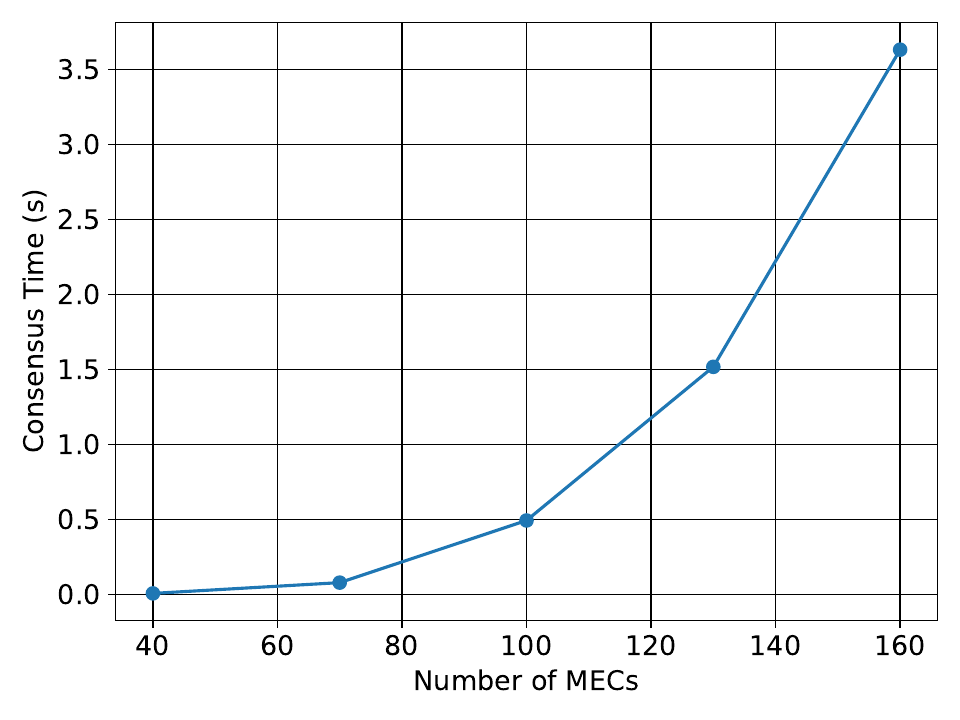}
\end{adjustbox}
\caption{Consensus Time (s)}
\label{fig:time}
\end{subfigure}
\caption{Performance of the blockchain system}
\label{fig:bc}
\end{figure*}

\begin{table}[!ht]
\centering
\caption{The detailed results of the blockchain system evaluation}
\label{tab:transconsensus}

\begin{tabular}{l|c c c c c}

\toprule[1pt]
\multirow{2}{*}{}                & \multicolumn{5}{c}{The number of MEC nodes} \\ \cline{2-6} 
                                 & 40             & 70             & 100 & 120 & 160            \\ \hline
Nb. transactions per block & 8 & 14 & 20 &  26 & 32            \\ \hline
Consensus time (s)             & 0.007 & 0.079 & 0.494 & 1.518 & 3.632         \\ \hline
Message Overhead              & 24& 84& 180& 312& 480          \\ 
\toprule[1pt]
\end{tabular}

\end{table}

Overall, the model's performance is quite satisfactory. The results in Table~\ref{tab:0-day} indicate that the effectiveness of the AC model can vary depending on the type of attack. However, it can identify most attacks with a high level of accuracy. Specifically, in the 5G-NIDD dataset, the model accurately detected 7 out of 8 attacks with an accuracy exceeding 85\%. Moreover, in the VDoS dataset, the model accurately detected 2 out of 3 attacks with an accuracy exceeding 93\%. It can be observed that the model has superior performance against scan-type attacks.

The model's limited accuracy in identifying Slowrate DoS attacks can be attributed to its resemblance to HTTP flood attack, which makes distinguishing between the two challenging.  It is crucial to note that closed-set supervised methods have also yielded similar misclassifications between HTTP flood and Slowrate DoS attack flows \cite{5GNIDD}. Despite the differences between the two attacks regarding speed and packet volume, they exhibit some similarities. In HTTP Flood attacks, a massive influx of HTTP requests overwhelms the server, whereas in Slowloris (first scenario of slowrate DoS \cite{5GNIDD}) attack, the attacker maintains multiple connections for extended periods, leading to resource depletion on the server.

The results of our experiment indicate that, on average, the detection rate for 60\% of attacks is 4.09\% better in the IID setting than in the Non-IID setting. In addition, the accuracy and AUROC are 3.5\% and 6.94\% better, respectively, in the IID setting. However, the TNR for ICMP flood and TCPConnectScan attacks are similar in both IID and Non-IID settings. Interestingly, the Non-IID setting outperforms the IID setting in the case of SYNFlood and UDPScan attacks. Overall, the AC model demonstrates promising performance in both Non-IID and IID settings. The Non-IID configuration is more challenging but, simultaneously, more representative of the real-world scenario where participating MECs (SOCs) may not have the same distribution of attack samples.

\subsection{Blockchain system's performance}
This section evaluates the performance of the blockchain system proposed for sharing model updates. In this experiment, we consider implementing this framework in Luxembourg as a case study. To comprehensively assess the system's performance, we plan to deploy a certain number of MEC servers, ranging from 40 to 160, at key base stations across each of Luxembourg's 12 administrative cantons. Considering the nature of our consortium blockchain, which is comprised of MEC nodes, the experiment includes varying the number of these nodes to examine scalability. We have configured the number of validators to be 10\% of the MEC nodes, while ensuring compliance with the PoA requirement of having $3f+1$ validators, where $f$ signifies the maximum number of potentially faulty MECs.

\begin{figure*}
    \centering
    \begin{subfigure}{.24\linewidth}
        \centering
        \includegraphics[width=\linewidth]{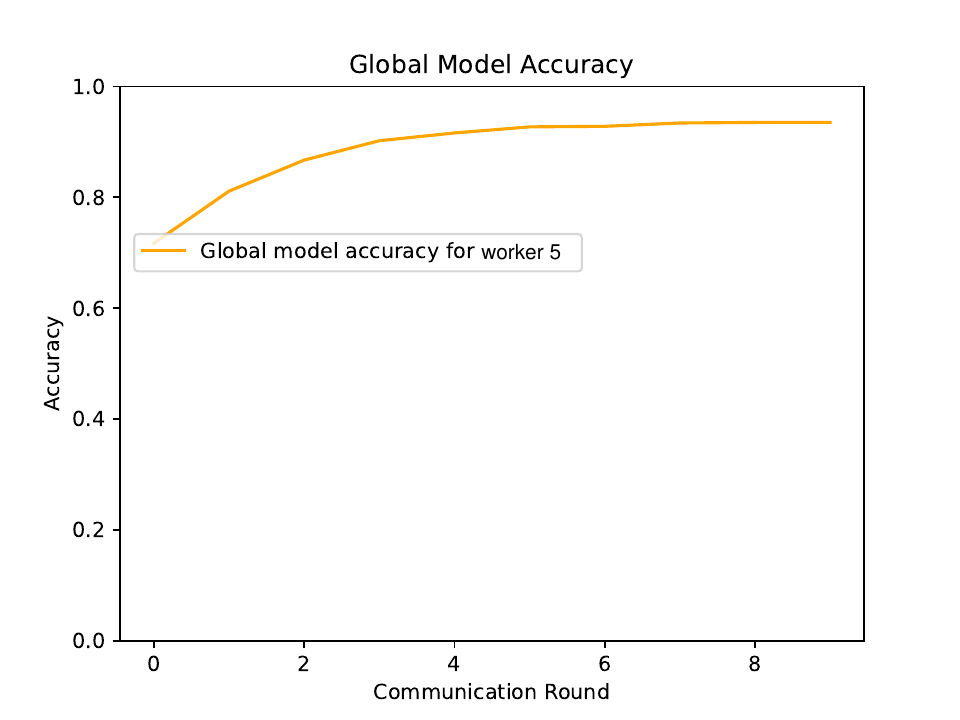} % Mettez à jour le chemin de votre image
        \caption{0\% malicious MW}
        \label{fig9a}
    \end{subfigure}
    \hfill
    \begin{subfigure}{.24\linewidth}
        \centering
        \includegraphics[width=\linewidth]{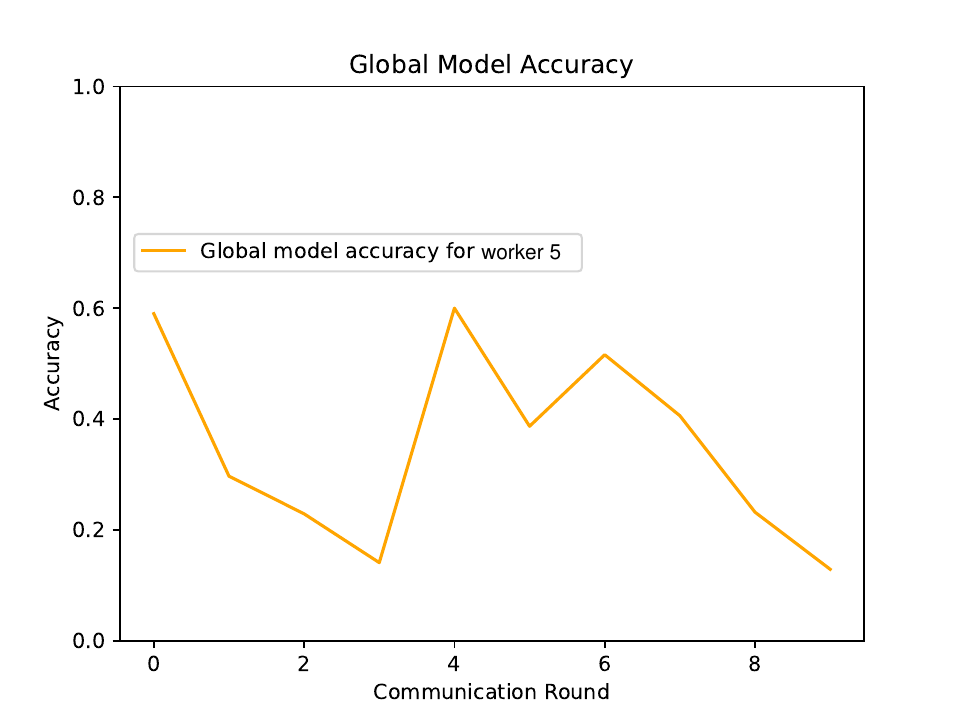} % Mettez à jour le chemin de votre image
        \caption{30\% MW \& PoA disabled}
        \label{fig9c} 
    \end{subfigure}
    \hfill
    \begin{subfigure}{.24\linewidth}
        \centering
        \includegraphics[width=\linewidth]{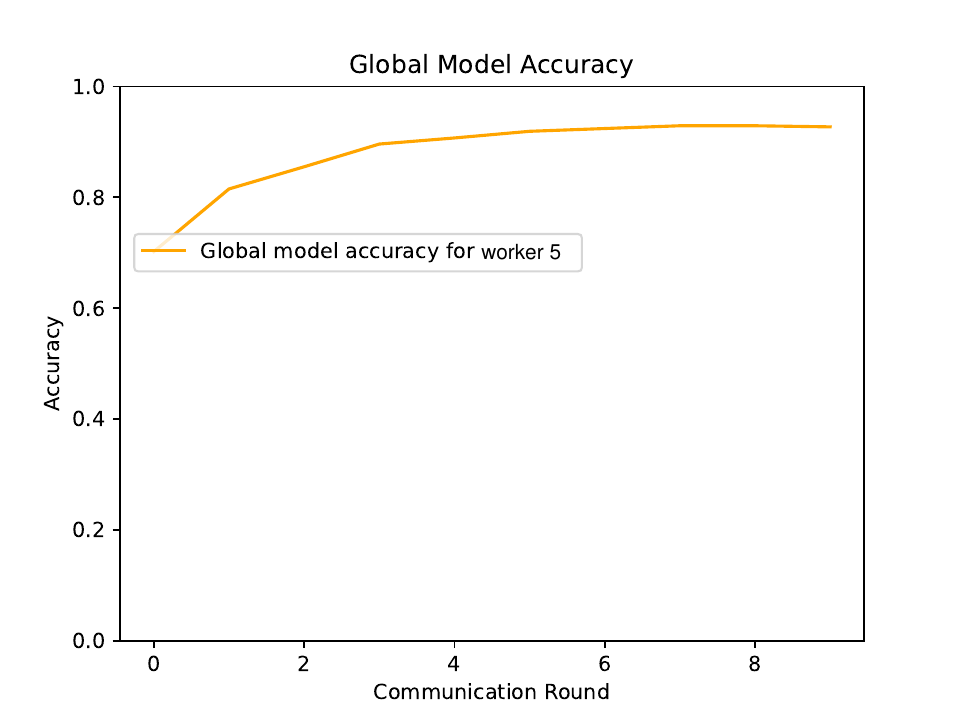} % Mettez à jour le chemin de votre image
        \caption{30\% MW \& PoA enabled}
        \label{fig10a}
    \end{subfigure}
    \hfill
    \begin{subfigure}{.24\linewidth}
        \centering
        \includegraphics[width=\linewidth]{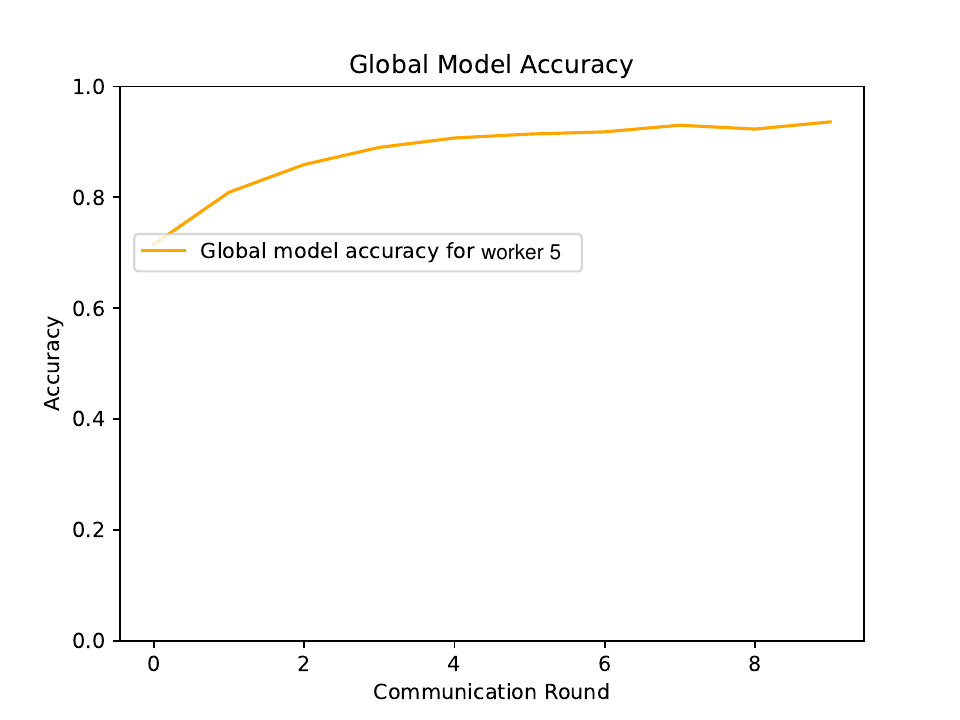} % Mettez à jour le chemin de votre image
        \caption{50\% MW \& PoA enabled}
        \label{fig10c}  
    \end{subfigure}
    \caption{Impact analysis of model poisoning attack on Zero-X Accuracy}
    \label{fig:modShare}
\end{figure*}

This evaluation was conducted on a machine equipped with a CPU (Intel i9 2.6 GHz) and 32 GB of RAM. The performance of the blockchain system is illustrated in Figure~\ref{fig:bc}, which displays metrics such as the message overhead, the number of transactions per block, and the consensus time. As anticipated, an increase in the number of MEC nodes leads to a corresponding rise in message overhead, the number of transactions per block, and consensus time, as observed in Figure \ref{fig:bc}. It is important to note, however, that despite these increases, both the consensus time and message overhead (shown in Figures \ref{fig:time} and \ref{fig:overhead}, respectively) continue to remain within reasonable limits. The architecture of the system demonstrates its ability to scale, as shown by the moderate increase in consensus time, even when the number of nodes quadruples. From Figure \ref{fig:time}, we can see that deploying 10 MEC nodes in each canton keeps the consensus time below 2 seconds, with transaction counts steady at 26, and message overhead stable at 180. Table~\ref{tab:transconsensus} presents a detailed overview of blockchain performance. 

Since CAVs operate as passive nodes, their increasing numbers do not influence the transaction volume per block due to the MEC nodes' role in aggregating AD model updates from the CAVs. Consequently, the addition of CAVs also does not impact consensus time, as they are not participants in the consensus mechanism. Given that each MEC is capable of serving an average of 500 participating CAVs, our system is projected to sustain its performance levels even with a network comprising approximately 120 MECs and 60,000 CAVs. The system's design ensures that the expansion in node count does not detract from blockchain performance, which affirms the system's scalability.

The robustness tests on the PoA mechanism under compromised MEC nodes reveal its effectiveness against poisoning attacks on the AC model. Figures \ref{fig9a} and \ref{fig9c} compare model accuracy without and with 30\% compromised workers, demonstrating a convergence failure without PoA. Upon enabling PoA, configured such that workers accumulating a negative ACC\_gain over two consecutive rounds are deemed unreliable and thus excluded from further training and consensus participation, the resilience of the system is evident. Figure \ref{fig10a} depicts the accuracy after enabling PoA, which confirms that our system effectively filtered out malicious nodes while maintaining performance nearly identical to that in a scenario free from such workers. Even with an increased presence of malicious workers to 50\%, Figure \ref{fig10c} shows the model still converges with comparable accuracy. We can, therefore, conclude that the solution is both effective and robust, even when faced with a model poisoning attack involving multiple nodes.

%When integrating blockchain with federated learning for attack detection, it is important to choose the right blockchain type and consensus protocol, as they affect scalability, latency, complexity, and cost. Public blockchains, whether PoW or PoS, have distinct integration challenges with federated learning. PoW struggles with limited scalability and high latency, leading to high costs and complexity, while PoS, despite better scalability and latency, faces public system constraints and complex token management. Private blockchains offer excellent scalability and low latency, advantageous for federated learning, but may conflict with its decentralization and privacy principles. Consortium blockchains with PoW or PoS balance these aspects but face latency issues (PoW) and governance complexities (PoS). BFT (Byzantine Fault Tolerance) consensus enhances security and reliability in federated learning, effectively managing malicious or faulty nodes, though it faces scalability and complexity challenges in large networks. Zero-X implements a consortium blockchain utilizing a BFT consensus mechanism with elected validators and role switching prolicy. This configuration optimizes scalability and enhances resilience against malicious behaviors, particularly during the verification of federated learning updates. This approach slightly compromises decentralization for improved scalability, latency, and security.

\subsection{Security Analysis}
This section provides a comprehensive security analysis of our proposed framework, addressing the mitigation of three critical threat vectors (discussed in section \ref{sec:threat}): Inter-vehicular attacks, attacks targeting MEC infrastructure, and adversarial attacks against the Zero-X framework.

\begin{itemize}
%\item \textbf{Inter-Vehicular Attacks}
%Single or multiple colluding malicious CAVs launch flooding attacks on a target CAV, hindering its ability to receive and process legitimate traffic. Zero-X's `Attack Detector' employs reconstruction error as an anomaly score to identify malicious flows in inter-CAV communications. Our study tested three types of DoS attacks in both N-day and 0-day scenarios. The effectiveness of Zero-X in these scenarios is demonstrated in Tables \ref{tab:N-day} and \ref{tab:0-day}, supporting our hypothesis regarding traffic pattern disruptions from such attacks. Once detected, packets sent by malicious CAVs will be systematically rejected

%Significantly, Zero-X quickly detects attacks within 1 second, achieving an F1-score over 88\%, and maintains efficiency in a short TW, as discussed in section~\ref{sec:tw}.

%\item \textbf{Attacks against the MEC infrastructure}
%In light of the MEC's crucial role in the IoV network, the threat of a compromised CAV botnet targeting the MEC system is a significant concern. The attack typically starts with a scan of the MEC server, followed by flood-type DDoS attacks through open ports. Deploying Zero-X at the MEC level, with its anomaly-scoring 'Attack Detector', has proven effective in mitigating such threats. Tests against five types of DoS attacks in both N-day and 0-day scenarios confirm Zero-X's efficacy, as indicated in Tables \ref{tab:N-day} and \ref{tab:0-day}, supporting our hypothesis that compromised CAVs exhibit distinguishable traffic patterns.

\item \textit{Inter-Vehicular and MEC Infrastructure attacks}: Single or multiple colluding malicious CAVs can launch flooding attacks on a target CAV, a scenario known as Inter-Vehicular Attacks. These attacks impair the target CAV's ability to process legitimate traffic. For a greater impact, these malicious CAVs, including compromised CAVs part of an IoV botnet, may also target the MEC infrastructure. In this case, the CAVs begin their attack by scanning the MEC server, followed by executing flood-type DDoS attacks through open ports. For both these attack vectors, Zero-X's 'Attack Detector' plays a pivotal role. It employs reconstruction error as an anomaly score to identify malicious traffic flows, whether in inter-CAV communications or targeting the MEC system. Our study conducted tests on three types of DoS attacks in inter-vehicular scenarios and five types in the context of MEC infrastructure, in both N-day and 0-day scenarios. The effectiveness of Zero-X is consistently demonstrated in these tests, as shown in Tables \ref{tab:N-day} and \ref{tab:0-day}. These results substantiate our hypothesis that both inter-vehicular and MEC-targeted attacks by compromised CAVs can be detected through distinct traffic pattern disruptions. Once identified, packets from malicious CAVs are systematically rejected, ensuring the integrity and security of the IoV network.

\color{black}

%\item \textit{Poisoning attack}: To prevent malicious CAVs or MECs from poising the training of the AD and AC global models by uploading poor-quality model parameters \cite{chen2021robust}, the proposed framework implements a model validation mechanism based on model accuracy gain. The MEC node first checks the AD model updates $AD_{Up}$ on $B_{Test}$, and a model is considered valid if the gap between $l(w^{t})$ and $l(w^{t-1})$ falls within a certain range $\delta$ (refer to section \ref{sec:poa}). If the update does not meet the validation criteria, it is rejected. To protect the training of the AC model, the framework uses a the PoA consensus mechanism that excludes MEC nodes with a negative accumulated \textit{ACC_gain} over a specified number of rounds from participating in subsequent training and consensus rounds. This measure safeguards the training process against potential model poisoning attacks that may arise from a compromised MEC. Furthermore, the role-switching policy of PoA ensures that validators are re-selected for each new round, minimizing the likelihood of a compromised MEC being repeatedly assigned to a validator role.
\item \textit{Poisoning attack}: To secure the training of the AD and AC global models against poisoning by malicious CAVs or MECs uploading substandard model parameters (see \cite{chen2021robust}), our framework incorporates a model validation mechanism that focuses on model accuracy gain. The process starts with the MEC node examining the AD model updates ($AD_{Up}$) against the test batch $B_{Test}$. A model update is deemed valid only if the difference in loss function values between $l(w^{t})$ and $l(w^{t-1})$ remains within a predefined threshold $\delta$ (detailed in section \ref{sec:poa}). Updates failing to satisfy this criterion are discarded. For the AC model's integrity, the framework employs a PoA consensus mechanism. It systematically excludes MEC nodes demonstrating a consistent negative \textit{ACC_gain} over several rounds from future training and consensus activities. This proactive approach effectively prevents model poisoning that could stem from a compromised MEC node. Additionally, the PoA's role-switching policy ensures periodic re-selection of validators for each training round, significantly reducing the risk of a compromised MEC node being repeatedly chosen as a validator.

\item \textit{Inference attack}: An adversary might attempt to infer raw data from a $i$-CAV using its uploaded weight and gradient information \cite{zhao2022practical} . To preserve privacy, Zero-X implements differential privacy in local updates. Upon completing local training, $CAV_{i}$ adds Gaussian noise to the trained parameters $w_{i}^{t}$. The gradient $w^{t}$ is then clipped: $w^{t+1} \leftarrow w^{t+1}/ \max (1, \frac{ ||w^{t+1}||}{C})$, where $C$ is the gradient norm bound. Subsequently, additive noise $n_{i}^{t}$ is incorporated into $w^{t}$. The resulting noised updates $\widetilde{w}{i}^{t}$ are then transmitted to the adjacent MEC node. 
\color{black}
\end{itemize}
%Recent research \cite{zhao2022practical} has highlighted that FL is vulnerable to inference attacks. Despite containing only weight and gradient information about private data, local models can still reveal sensitive information, such as labels, memberships, and properties, allowing adversaries to reconstruct the private data. To counter this threat, the Zero-X leverages a differential privacy scheme to safeguard the privacy of CAV's native data from honest-but-curious MEC nodes. This is achieved by perturbing the AD local model updates by injecting noise into the gradient information before uploading to the MEC nodes.

\begin{figure*}[htbp] % Utilisez figure* pour une figure sur deux colonnes
  \centering
    \begin{subfigure}[b]{.24\textwidth}
    \centering
    \begin{adjustbox}{width=\linewidth}
\begin{tikzpicture}
  \begin{axis}[
    ybar,
    %ylabel={F1-Score (\%)}, % Uncomment if needed
    symbolic x coords={0-day, N-day},
    xtick=data,
    ymajorgrids=true,
    grid style=dashed,
    legend style={at={(0.5,-0.15)},anchor=north,legend columns=-1},
    ymin=70, % Adjust the minimum y value if needed
    enlarge x limits=0.25,
    bar width=15pt, % Adjust the bar width if needed
  ]
    \addplot coordinates {(0-day,98.36) (N-day,93.39)};
    \addplot[fill=orange!30!white] coordinates {(0-day,75.943) (N-day,99.8)};
    \legend{Zero-X, MTH-IDS \cite{yang2021mth}}
  \end{axis}
\end{tikzpicture}
    \end{adjustbox}
    \caption{CIC-IDS-2017}
    \label{fig:comMth}
  \end{subfigure}
  \begin{subfigure}[b]{.24\textwidth} % Alignement en bas
    \centering
    \begin{adjustbox}{width=\linewidth}
    \begin{tikzpicture}
      \begin{axis}[
        ybar,
        symbolic x coords={HTTP, ICMP, SYN, UDP, Slow},
        xtick=data,
        ymajorgrids=true,
        grid style=dashed,
        legend style={at={(0.5,-0.15)},anchor=north,legend columns=-1},
        ymin=90,
        enlarge x limits=0.15,
        bar width=7pt,
      ]
        \addplot coordinates {(HTTP,100) (ICMP,100) (SYN,100) (UDP,100) (Slow,96.9)};
        \addplot coordinates {(HTTP,99.05) (ICMP,99.97) (SYN,99.44) (UDP,99.82) (Slow,98.11)};
        \legend{Zero-X, 5G-NIDD\cite{5GNIDD}}
      \end{axis}
    \end{tikzpicture}
    \end{adjustbox}
    \caption{5G-NIDD DoS Attacks}
    \label{fig:com5gDos}
  \end{subfigure}%
  \begin{subfigure}[b]{.24\textwidth}
    \centering
    \begin{adjustbox}{width=\linewidth}
    \begin{tikzpicture}
      \begin{axis}[
        ybar,
        %ylabel={F1-Score (\%)}, % Décommentez si nécessaire
        symbolic x coords={SYN, TCP, UDP},
        xtick=data,
        ymajorgrids=true,
        grid style=dashed,
        legend style={at={(0.5,-0.15)},anchor=north,legend columns=-1},
        ymin=90,
        enlarge x limits=0.15,
        bar width=7pt,
      ]
        \addplot coordinates {(SYN,99.75) (TCP,99.46) (UDP,99.84)};
        \addplot coordinates {(SYN,99.71) (TCP,98.45) (UDP,96.27)};
        \legend{Zero-X, 5G-NIDD \cite{5GNIDD} }
      \end{axis}
    \end{tikzpicture}
    \end{adjustbox}
    \caption{5G-NIDD Scan Attacks}
    \label{fig:com5gScan}
  \end{subfigure}%
  \begin{subfigure}[b]{.24\textwidth}
    \centering
    \begin{adjustbox}{width=\linewidth}
    \begin{tikzpicture}
      \begin{axis}[
        ybar,
        %ylabel={F1-Score (\%)}, % Décommentez si nécessaire
        symbolic x coords={SYN,UDP,Slow},
        xtick=data,
        ymajorgrids=true,
        grid style=dashed,
        legend style={at={(0.5,-0.15)},anchor=north,legend columns=-1},
        ymin=0,
        enlarge x limits=0.15,
        bar width=7pt,
      ]
        \addplot coordinates {(SYN,100) (UDP,100) (Slow,100)};
        \addplot[fill=yellow] coordinates {(SYN,90.05) (UDP,89.90) (Slow,0)};
        \addplot coordinates {(SYN,100) (UDP,100) (Slow,99.95)};
        \legend{Zero-X, FLZAD \cite{korba2023federated}, VDoS \cite{VDoS}}
      \end{axis}
    \end{tikzpicture}
    \end{adjustbox}
    \caption{VDoS Attacks}
    \label{fig:comVDOS}
  \end{subfigure}
  
  \caption{Comparative Performance Analysis of Zero-X with Related Works}
  \label{fig:comparisons}
\end{figure*}
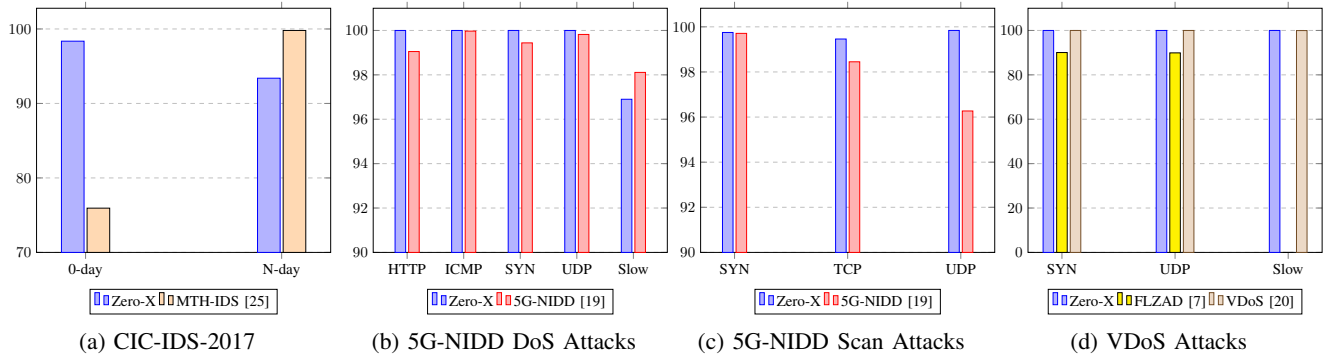

\begin{table*}
\caption{Comparison between Zero-X and other related works}
\centering
%\begin{adjustbox}{width=0.5\textwidth}
\begin{tabular}{@{}llllllllllll@{}}
%Content\footnote{footnote text}
\toprule
Features & ~\cite{korba2023federated} & \cite{5GNIDD} & \cite{abdel2021federated} & \cite{VDoS} & \cite{boualouache2023demand} & \cite{selamnia2022edge} & \cite{boualouache2022federated} & \cite{agrawal2022novelads} & \cite{jeong2023x} & \cite{yang2021mth} & \textbf{Zero-X} \\ \midrule
0-day   detection & \checkmark &  \xmark & \xmark & \xmark & \xmark & \xmark & \xmark & \checkmark & \checkmark & \checkmark & \checkmark \\
N-day   identificatin & \xmark & \checkmark & \checkmark & \checkmark & \checkmark & \checkmark & \checkmark & \xmark & \xmark & \checkmark & \checkmark \\
Federated   learning & \checkmark & \xmark & \checkmark & \xmark & \checkmark & \checkmark & \checkmark & \xmark & \xmark & \xmark & \checkmark \\
IID & \checkmark & \xmark & \checkmark & \xmark & \checkmark & \checkmark & \checkmark & \xmark & \xmark & \xmark & \checkmark \\
Non-IID & \xmark & \xmark & \xmark & \xmark & \xmark & \xmark & \xmark & \xmark & \xmark & \xmark & \checkmark \\
Decentralized Training & \xmark & \xmark & \checkmark & \xmark & \checkmark & \xmark & \xmark & \xmark & \xmark & \xmark & \checkmark \\
Fault   tolerance & \xmark & \xmark & \checkmark & \xmark & \checkmark & \xmark & \xmark & \xmark & \xmark & \xmark & \checkmark \\
Privacy preserving & \xmark & \xmark & \xmark & \xmark & \xmark & \xmark & \xmark & \xmark & \xmark & \xmark & \checkmark \\
Real vehicular data & \checkmark & \xmark & \xmark & \checkmark & \xmark & \xmark & \xmark & \checkmark & \checkmark & \checkmark & \checkmark \\
5G network traces & \xmark & \checkmark & \xmark & \xmark & \checkmark & \xmark & \xmark & \xmark & \xmark & \xmark & \checkmark \\ \bottomrule
%\multicolumn{12}{l}{\small \checkmark: Yes, \xmark, No} \\
\end{tabular}
%\end{adjustbox}
\label{tab:compSota}
\end{table*}

\subsection{Discussion and comparison}
In the current state-of-the-art, only a few studies \cite{korba2023federated, khan2022enhanced, agrawal2022novelads, yang2021mth} have addressed the detection of zero-day attacks in the IoV. Agrawal et al. \cite{agrawal2022novelads} primarily focused on intra-vehicular networks, whereas Khan et al. \cite{khan2022enhanced, yang2021mth} extended their research to both intra and inter-vehicular networks. Only Yang et al. \cite{yang2021mth} conducted experiments specifically focused on zero-day scenarios. We conducted a comparative analysis with their work using the CIC-IDS-2017 dataset \cite{sharafaldin2018toward}. As demonstrated in Figure \ref{fig:comMth}, the performance difference between MTH-IDS and Zero-X in detecting N-Day attacks is relatively small, with MTH-IDS achieving a 99.8\% detection rate, compared to 93.39\% for Zero-X. In the context of 0-Day attacks, however, Zero-X significantly outperforms MTH-IDS, achieving a 98.36\% detection rate versus 75.943\% for MTH-IDS. Notably, MTH-IDS employs centralized learning, which raises privacy concerns in the IoV due to the necessity of data centralization. We further compared Zero-X with three recent studies that utilized the same datasets \cite{5GNIDD,VDoS} as those used in our experimentation. Figure \ref{fig:com5gDos} shows Zero-X outperforming Schan et al.'s method \cite{5GNIDD} in DDoS attack detection, except for SlowrateDoS. As demonstrated in Figure \ref{fig:com5gScan}, Zero-X also excels in detecting Scan-type attacks. Figure \ref{fig:comVDOS} reveals Zero-X's superiority over Rahal et al. \cite{VDoS} and \cite{korba2023federated} in inter-vehicular attack detection. Zero-X addresses the limitations of \cite{VDoS}'s inability to detect zero-day attacks and \cite{korba2023federated}'s inability to recognize N-day attacks.

The practicality and real-world potential of the Zero-X framework are demonstrated through evaluations conducted on both a realistic vehicular traffic testbed \cite{VDoS} and real traffic attacks on MEC infrastructure \cite{5GNIDD}. Both datasets contain diverse attacks mainly categorized as DoS and Scan. Their similar characteristics lead to resembling network behaviors, complicating the detection of zero-day attacks. These new patterns may mimic known attacks, challenging the classification model's ability to differentiate them. Table~\ref{tab:compSota} compares the features of the Zero-X framework with other recent studies in various key aspects. Notably, Zero-X stands out by effectively detecting 0-day and recognizing N-day attacks, a capability shared with MTH-IDS \cite{yang2021mth}. However, it goes further by prioritizing data privacy through federated learning. Additionally, Zero-X leverages blockchain technology to decentralize and distribute the coordination of federated training. This not only enhances robustness against adversarial attacks but also minimizes the risks associated with single points of failure and potential data tampering.

\color{black}
%Qualitative

% \usepackage{booktabs}
%\makesavenoteenv{tabular}

%\textcolor{blue}{In integrating blockchain with FL, choosing the suitable blockchain type and consensus protocol is crucial due to their impact on scalability, latency, complexity, and cost. Public blockchains (PoW or PoS) present challenges like limited scalability, high latency, and token management complexities. Private blockchains provide good scalability and low latency but may compromise decentralization and privacy. Consortium blockchains offer a balance but deal with latency (PoW) and governance complexities (PoS). BFT consensus improves security in FL but can struggle with scalability and complexity in larger networks. Zero-X implements a consortium blockchain utilizing PoA, a BFT consensus mechanism with elected validators and role switching prolicy. This configuration optimizes scalability and enhances resilience against malicious behaviors through the validation of FL updates. This approach slightly compromises decentralization for improved scalability, latency, and security.}

%======================
\section{Conclusion} \label{CON}

%% New  (Edited )
This paper proposed Zero-X, a novel security framework designed to detect both 0-day and N-day attacks in the realm of the IoV. The framework uses a blockchain-enabled federated training approach to build detection and identification models that enable secure collaboration and trust between CAV and MEC nodes. We empirically evaluate the proposed framework and compare its performance with existing IDS using recent publicly available datasets. The results show that the Zero-X framework outperforms existing IDS, demonstrating its superiority in detecting a wide range of cyberattacks. For our future work, we plan to develop a security framework with the ability to detect zero-day attacks, as well as evolve its detection capabilities. This system will incrementally update the detection model to convert newly discovered "unknown" attacks into "known" attacks. Another direction to extend this work is to integrate an intrusion response mechanism where multiple reinforcement learning agents are running on a set of vehicular edge nodes and learn to mitigate newly discovered attacks.

\section*{Acknowledgments}
This work was supported by the 5G-INSIGHT bilateral project (ID: 14891397) / (ANR-20-CE25-0015-16), funded by the Luxembourg National Research Fund (FNR), by the French National Research Agency (ANR), and by the FEDER MISMAR, Région Nouvelle-Aquitaine B4IoT.

%\begin{thebibliography}{1}
\bibliographystyle{IEEEtran}
\bibliography{ref.bib}

\begin{thebibliography}{10}
\providecommand{\url}[1]{#1}
\csname url@samestyle\endcsname
\providecommand{\newblock}{\relax}
\providecommand{\bibinfo}[2]{#2}
\providecommand{\BIBentrySTDinterwordspacing}{\spaceskip=0pt\relax}
\providecommand{\BIBentryALTinterwordstretchfactor}{4}
\providecommand{\BIBentryALTinterwordspacing}{\spaceskip=\fontdimen2\font plus
\BIBentryALTinterwordstretchfactor\fontdimen3\font minus \fontdimen4\font\relax}
\providecommand{\BIBforeignlanguage}[2]{{%
\expandafter\ifx\csname l@#1\endcsname\relax
\typeout{** WARNING: IEEEtran.bst: No hyphenation pattern has been}%
\typeout{** loaded for the language `#1'. Using the pattern for}%
\typeout{** the default language instead.}%
\else
\language=\csname l@#1\endcsname
\fi
#2}}
\providecommand{\BIBdecl}{\relax}
\BIBdecl

\bibitem{guo2022review}
Y.~Guo, ``A review of machine learning-based zero-day attack detection: Challenges and future directions,'' \emph{Computer Communications}, 2022.

\bibitem{turgeman_2022}
\BIBentryALTinterwordspacing
N.~Turgeman, ``Infographic: Top real-world threats facing connected cars and fleets,'' Sep 2022. [Online]. Available: \url{https://upstream.auto/blog/infographic-top-real-world-threats-facing-connected-cars-fleets/}
\BIBentrySTDinterwordspacing

\bibitem{boualouache2022survey}
A.~Boualouache and T.~Engel, ``{A Survey on Machine Learning-based Misbehavior Detection Systems for 5G and Beyond Vehicular Networks},'' \emph{arXiv preprint arXiv:2201.10500}, 2022.

\bibitem{abdel2021federated}
M.~Abdel-Basset, N.~Moustafa, H.~Hawash, I.~Razzak, K.~M. Sallam, and O.~M. Elkomy, ``Federated intrusion detection in blockchain-based smart transportation systems,'' \emph{IEEE Transactions on Intelligent Transportation Systems}, vol.~23, no.~3, pp. 2523--2537, 2021.

\bibitem{rahal2022antibotv}
R.~Rahal, A.~Amara~Korba, N.~Ghoualmi-Zine, Y.~Challal, and M.~Y. Ghamri-Doudane, ``Antibotv: A multilevel behaviour-based framework for botnets detection in vehicular networks,'' \emph{Journal of Network and Systems Management}, vol.~30, pp. 1--40, 2022.

\bibitem{boualouache2022federated}
A.~Boualouache and T.~Engel, ``Federated learning-based inter-slice attack detection for 5g-v2x sliced networks,'' in \emph{2022 IEEE 96th Vehicular Technology Conference (VTC2022-Fall)}.\hskip 1em plus 0.5em minus 0.4em\relax IEEE, 2022, pp. 1--6.

\bibitem{korba2023federated}
A.~A. Korba, A.~Boualouache, B.~Brik, R.~Rahal, Y.~Ghamri-Doudane, and S.~M. Senouci, ``Federated learning for zero-day attack detection in 5g and beyond v2x networks,'' in \emph{AlgoTel 2023-25{\`e}mes Rencontres Francophones sur les Aspects Algorithmiques des T{\'e}l{\'e}communications}, 2023.

\bibitem{jeong2023x}
S.~Jeong, S.~Lee, H.~Lee, and H.~K. Kim, ``X-canids: Signal-aware explainable intrusion detection system for controller area network-based in-vehicle network,'' \emph{arXiv preprint arXiv:2303.12278}, 2023.

\bibitem{ashraf2020novel}
J.~Ashraf, A.~D. Bakhshi, N.~Moustafa, H.~Khurshid, A.~Javed, and A.~Beheshti, ``Novel deep learning-enabled lstm autoencoder architecture for discovering anomalous events from intelligent transportation systems,'' \emph{IEEE Transactions on Intelligent Transportation Systems}, vol.~22, no.~7, pp. 4507--4518, 2020.

\bibitem{geng2020recent}
C.~Geng, S.-j. Huang, and S.~Chen, ``Recent advances in open set recognition: A survey,'' \emph{IEEE transactions on pattern analysis and machine intelligence}, vol.~43, no.~10, pp. 3614--3631, 2020.

\bibitem{kong2021opengan}
S.~Kong and D.~Ramanan, ``Opengan: Open-set recognition via open data generation,'' in \emph{Proceedings of the IEEE/CVF International Conference on Computer Vision}, 2021, pp. 813--822.

\bibitem{neal2018open}
L.~Neal, M.~Olson, X.~Fern, W.-K. Wong, and F.~Li, ``Open set learning with counterfactual images,'' in \emph{Proceedings of the European Conference on Computer Vision (ECCV)}, 2018, pp. 613--628.

\bibitem{oza2019c2ae}
P.~Oza and V.~M. Patel, ``C2ae: Class conditioned auto-encoder for open-set recognition,'' in \emph{Proceedings of the IEEE/CVF Conference on Computer Vision and Pattern Recognition}, 2019, pp. 2307--2316.

\bibitem{zhang2022unknown}
Z.~Zhang, Y.~Zhang, J.~Niu, and D.~Guo, ``Unknown network attack detection based on open-set recognition and active learning in drone network,'' \emph{Transactions on Emerging Telecommunications Technologies}, vol.~33, no.~10, p. e4212, 2022.

\bibitem{9643172}
S.~Xu, L.~Li, H.~Yang, and J.~Tang, ``Kcc method: Unknown intrusion detection based on open set recognition,'' in \emph{2021 IEEE 33rd International Conference on Tools with Artificial Intelligence (ICTAI)}, 2021, pp. 1343--1347.

\bibitem{uprety2021privacy}
A.~Uprety, D.~B. Rawat, and J.~Li, ``{Privacy Preserving Misbehavior Detection in IoV using Federated Machine Learning},'' in \emph{2021 IEEE 18th Annual Consumer Communications \& Networking Conference (CCNC)}.\hskip 1em plus 0.5em minus 0.4em\relax IEEE, 2021, pp. 1--6.

\bibitem{liu2021blockchain}
H.~Liu, S.~Zhang, P.~Zhang, X.~Zhou, X.~Shao, G.~Pu, and Y.~Zhang, ``Blockchain and federated learning for collaborative intrusion detection in vehicular edge computing,'' \emph{IEEE Transactions on Vehicular Technology}, vol.~70, no.~6, pp. 6073--6084, 2021.

\bibitem{hbaieb2022federated}
A.~Hbaieb, S.~Ayed, and L.~Chaari, ``{Federated learning based IDS approach for the IoV},'' in \emph{Proceedings of the 17th International Conference on Availability, Reliability and Security}, 2022, pp. 1--6.

\bibitem{5GNIDD}
S.~Samarakoon, Y.~Siriwardhana, P.~Porambage, M.~Liyanage, S.-Y. Chang, J.~Kim, J.~Kim, and M.~Ylianttila, ``5g-nidd: A comprehensive network intrusion detection dataset generated over 5g wireless network,'' \emph{arXiv preprint arXiv:2212.01298}, 2022.

\bibitem{VDoS}
R.~Rahal, A.~Amara~Korba, and N.~Ghoualmi-Zine, ``Towards the development of realistic dos dataset for intelligent transportation systems,'' \emph{Wireless Personal Communications}, vol. 115, no.~2, pp. 1415--1444, 2020.

\bibitem{anbalagan2023iids}
S.~Anbalagan, G.~Raja, S.~Gurumoorthy, R.~D. Suresh, and K.~Dev, ``Iids: Intelligent intrusion detection system for sustainable development in autonomous vehicles,'' \emph{IEEE Transactions on Intelligent Transportation Systems}, 2023.

\bibitem{lai2023improved}
Q.~Lai, C.~Xiong, J.~Chen, W.~Wang, J.~Chen, T.~R. Gadekallu, M.~Cai, and X.~Hu, ``Improved transformer-based privacy-preserving architecture for intrusion detection in secure v2x communications,'' \emph{IEEE Transactions on Consumer Electronics}, 2023.

\bibitem{khan2022enhanced}
I.~A. Khan, N.~Moustafa, D.~Pi, W.~Haider, B.~Li, and A.~Jolfaei, ``An enhanced multi-stage deep learning framework for detecting malicious activities from autonomous vehicles,'' \emph{IEEE Transactions on Intelligent Transportation Systems}, vol.~23, no.~12, pp. 25\,469--25\,478, 2022.

\bibitem{agrawal2022novelads}
K.~Agrawal, T.~Alladi, A.~Agrawal, V.~Chamola, and A.~Benslimane, ``Novelads: A novel anomaly detection system for intra-vehicular networks,'' \emph{IEEE Transactions on Intelligent Transportation Systems}, vol.~23, no.~11, pp. 22\,596--22\,606, 2022.

\bibitem{yang2021mth}
L.~Yang, A.~Moubayed, and A.~Shami, ``Mth-ids: A multitiered hybrid intrusion detection system for internet of vehicles,'' \emph{IEEE Internet of Things Journal}, vol.~9, no.~1, pp. 616--632, 2022.

\bibitem{9757866}
H.~Sedjelmaci and N.~Ansari, ``On cooperative federated defense to secure multi-access edge computing,'' \emph{IEEE Consumer Electronics Magazine}, pp. 1--1, 2022.

\bibitem{antibot}
R.~Rahal, A.~Amara~Korba, N.~Ghoualmi-Zine, Y.~Challal, and M.~Y. Ghamri-Doudane, ``Antibotv: A multilevel behaviour-based framework for botnets detection in vehicular networks,'' \emph{Journal of Network and Systems Management}, vol.~30, no.~1, pp. 1--40, 2022.

\bibitem{chen2021robust}
H.~Chen, S.~A. Asif, J.~Park, C.-C. Shen, and M.~Bennis, ``Robust blockchained federated learning with model validation and proof-of-stake inspired consensus,'' \emph{arXiv preprint arXiv:2101.03300}, 2021.

\bibitem{zhao2022practical}
P.~Zhao, Z.~Cao, J.~Jiang, and F.~Gao, ``Practical private aggregation in federated learning against inference attack,'' \emph{IEEE Internet of Things Journal}, vol.~10, no.~1, pp. 318--329, 2022.

\bibitem{deepmcdd}
D.~Lee, S.~Yu, and H.~Yu, ``Multi-class data description for out-of-distribution detection,'' in \emph{Proceedings of the 26th ACM SIGKDD International Conference on Knowledge Discovery \& Data Mining}, 2020, pp. 1362--1370.

\bibitem{ae}
G.~E. Hinton and R.~Zemel, ``Autoencoders, minimum description length and helmholtz free energy,'' \emph{Advances in neural information processing systems}, vol.~6, 1993.

\bibitem{dbft}
Q.~Wang, J.~Yu, Z.~Peng, V.~C. Bui, S.~Chen, Y.~Ding, and Y.~Xiang, ``Security analysis on dbft protocol of neo,'' in \emph{Financial Cryptography and Data Security: 24th International Conference, FC 2020, Kota Kinabalu, Malaysia, February 10--14, 2020 Revised Selected Papers 24}.\hskip 1em plus 0.5em minus 0.4em\relax Springer, 2020, pp. 20--31.

\bibitem{cic}
``Cicflowmeter,'' \url{https://github.com/CanadianInstituteForCybersecurity/CI}, accessed: 2023-04-26.

\bibitem{FedAvg}
H.~B. McMahan, E.~Moore, D.~Ramage, S.~Hampson, and B.~A. y~Arcas, ``Communication-efficient learning of deep networks from decentralized data,'' in \emph{AISTATS}, 2017.

\bibitem{boualouache2023demand}
A.~Boualouache, B.~Brik, S.-M. Senouci, and T.~Engel, ``On-demand security framework for 5gb vehicular networks,'' \emph{IEEE Internet of Things Magazine}, 2023.

\bibitem{selamnia2022edge}
A.~Selamnia, B.~Brik, S.~M. Senouci, A.~Boualouache, and S.~Hossain, ``Edge computing-enabled intrusion detection for c-v2x networks using federated learning,'' in \emph{GLOBECOM 2022-2022 IEEE Global Communications Conference}.\hskip 1em plus 0.5em minus 0.4em\relax IEEE, 2022, pp. 2080--2085.

\bibitem{sharafaldin2018toward}
I.~Sharafaldin, A.~H. Lashkari, and A.~A. Ghorbani, ``Toward generating a new intrusion detection dataset and intrusion traffic characterization.'' \emph{ICISSp}, vol.~1, pp. 108--116, 2018.

\end{thebibliography}
 

%\bf{}\vspace{-33pt}
\begin{IEEEbiography}
[{\includegraphics[width=1in,height=1.25in,clip,keepaspectratio]{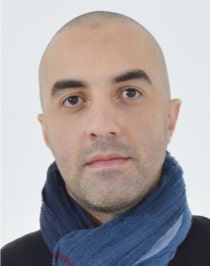}}]{Abdelaziz Amara korba}
is an associate professor in computer science. He received his Ph.D. in Computer Science from the University of Annaba in 2016, in a collaborative program with the University of La Rochelle, France. His research focuses on network security, machine learning (ML) applications in cybersecurity, intrusion detection systems, decentralized ML techniques, and the development of blockchain-based security solutions for the Internet of Things (IoT) and connected vehicles.
\end{IEEEbiography}

\begin{IEEEbiography}
[{\includegraphics[width=1in,height=1.25in,clip,keepaspectratio]{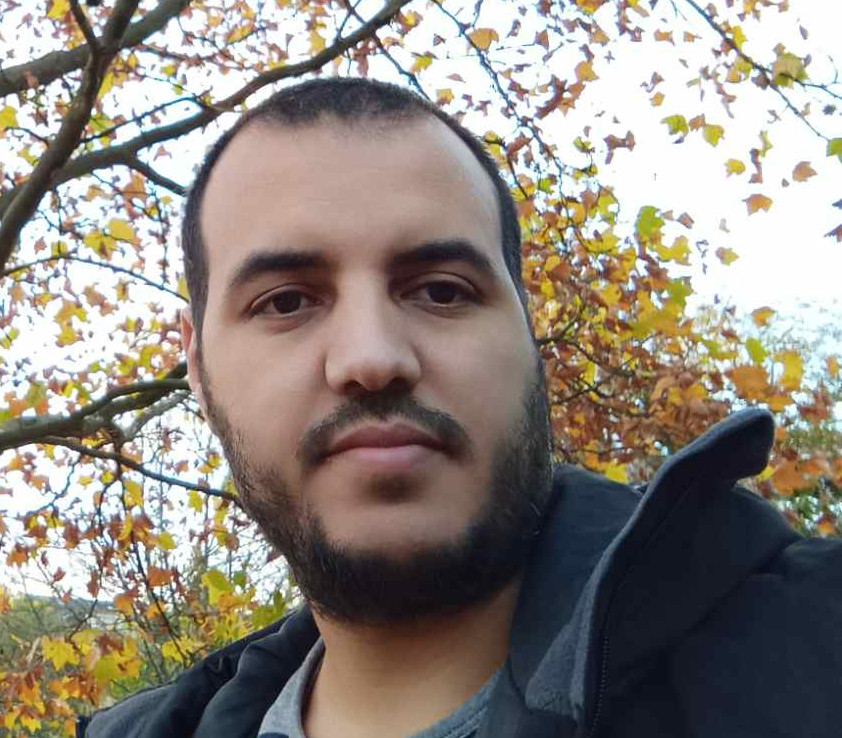}}]%
{Abdelwahab Boualouache} is a research associate at the Faculty of Science, Technology, and Medicine (FSTM), University of Luxembourg. He received a Ph.D. degree in computer science from USTHB University, Algiers, Algeria, in 2016. His current research interests include security and privacy in connected vehicles and privacy-preserving collaborative learning solutions for 5G and Blockchain-based solutions for decentralized systems
\end{IEEEbiography}

\begin{IEEEbiography}
[{\includegraphics[width=1in,height=1.25in,clip,keepaspectratio]{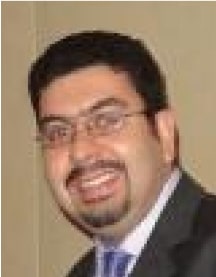}}]{Yacine Ghamri-Doudane}
is currently a Full Professor at the University of La Rochelle (ULR) in France and the director of its Laboratory of Informatics, Image and Interaction (L3i). His current research interests lays in the area of wireless networking and mobile computing with a current emphasis on topics related to the Internet of Things (IoT), Vehicular Networks as well as Digital Trust.
\end{IEEEbiography}

\end{document}